\documentclass[aps,prl,onecolumn]{revtex4}
\usepackage[latin1]{inputenc}
\usepackage{graphicx}%
\usepackage{amsmath}
\usepackage{latexsym}
\usepackage{amssymb}

\newcommand{\ud}{\,\mathrm{d}}

\begin{document}
%\bibliographystyle{elsart-num}
%\bibliographystyle{h-elsevier2}
%\bibliographystyle
%\bibliographystyle{elsart-num}
%\begin{frontmatter}

\title{X-ray scattering from stepped and kinked surfaces: an approach with the paracrystal model}

\author{Fr\'ed\'eric Leroy }
\email{leroy@crmcn.univ-mrs.fr}
\affiliation{Centre de Recherche en Mati\`ere Condens\'ee et NanoSciences, \\
CNRS - UPR 7281\\
Campus de Luminy Case 913, 13288 Marseille Cedex 09, France}

\author{R\'emi Lazzari }
\affiliation{Institut des NanoSciences de Paris \\ Universités
Pierre et
Marie Curie (Paris 6) et Denis Diderot (Paris 7), CNRS UMR 7588 \\
Campus de Boucicaut, 140 rue de Lourmel, 75015 Paris, France}

\author{Gilles Renaud }
%\email{grenaud@cea.fr}
%\email[Corresponding author :]{grenaud@cea.fr}
\affiliation{Commissariat à l'Energie Atomique,\\
D\'epartement de Recherche Fondamentale sur la Mati\`ere Condens\'ee, \\
Service de Physique des Mat\'eriaux et Microstructures,
Nanostructures et Rayonnement Synchrotron, \\17, Avenue des Martyrs,
F-38054 Grenoble, Cedex 9, France}

\begin{abstract}
A general formalism of X-ray scattering from different kinds of
surface morphologies is described. Based on a description of the
surface morphology at the atomic scale through the use of the
paracrystal model and discrete distributions of distances, the
scattered intensity by non-periodic surfaces is calculated over the
whole reciprocal space. In one dimension, the scattered intensity by
a vicinal surface, the two-level model, the N-level model, the
faceted surface and the rough surface are addressed. In two
dimensions, the previous results are generalized to the kinked
vicinal surface, the two-level vicinal surface and the step
meandering on a vicinal surface. The concept of Crystal Truncation
Rod is generalized considering also the truncation of a terrace by a
step (yielding a Terrace Truncation Rod) and a step by a kink
(yielding a Step Truncation Rod).

\end{abstract}

%\date{\today}
\maketitle

%\tableofcontents

% main text
%\begin{linenumbers}
\section{Introduction}

The control of the morphology and structure of surfaces is one key
point to elaborate high quality devices, since defects such as
steps, kinks or more generally surface roughness may drastically
influence their physical properties. Defects may also be used to
promote catalytic reactions or as preferential nucleation sites for
the self-assembling of nanostructures \cite{brune98a, leroy05}, e.
g. in view of ultra-high density recording. In this context, the
development of surface characterization tools is highly appreciated.
If scanning probe microscopy is the technique of choice to get local
real space information, statistical averages and \emph{in situ}
experiments are advantageously performed with diffraction
techniques. Numerous real time studies of thin films growth,
coarsening or roughening have been performed, for instance with
X-ray \cite{renaud01, braun04, renaud92}, Spot Profile Analysis Low
Energy Electron Diffraction (SPA-LEED) \cite{tegenkamp02, sotto00},
Helium Atom Scattering (HAS) \cite{farias98,
crottini97, ernst92} or Reflection High Energy Electron Diffraction (RHEED) \cite{dulot03} techniques. \\
In particular X-ray scattering have the advantage over the other
techniques of a weak scattering cross section which makes possible
the quantitative analysis of the diffraction patterns in the
framework of the kinematic approximation. The purpose of this paper
is to revisit the scattering theory by disordered stepped and kinked
surfaces in the context of the development of grazing incidence
X-ray studies in surface science. It extends elder works which
essentially dealt with Helium and electron diffraction.

The main problem in the calculation of the diffraction by disordered
surfaces is to have a statistical description of the morphology and
to be able to calculate the scattered intensity. If the morphology
is perfectly periodic, the scattered intensity is calculated in a
straightforward way. The structure factor of the unit cell which
repeats from the surface to the bulk, i.e. the Fourier transform of
its electronic density, has to be evaluated. Then the scattered
intensity is the modulus square of the sum of the amplitude
scattered by all the unit cells. In the special case of surface
diffraction, a semi-infinite sum has to be done in the direction
perpendicular to the surface. If one considers the simple case of a
cubic lattice, the intensity reads:

\begin{eqnarray}
I(q_{x},q_{y},q_{z}) &=& \mid \widetilde{F}_{cell}(q_{x},q_{y},q_{z})\mid^2\sum_{n_{1}=-\infty}^{+\infty}\sum_{n_{2}=-\infty}^{+\infty}\sum_{n_{3}=0}^{+\infty}e^{in_{1}q_{x}a}e^{in_{2}q_{y}b}e^{in_{3}q_{z}c} \nonumber \\
&=& \frac{\mid \widetilde{F}_{cell}(q_{x},q_{y},q_{z})\mid
^2}{(2\sin (q_{z}
c/2))^2}\sum_{n_{1}=-\infty}^{+\infty}\sum_{n_{2}=-\infty}^{+\infty}\delta
\left( q_{x}-n_{1}\frac{2\pi}{a}\right )\delta
\left(q_{y}-n_{2}\frac{2\pi}{b}\right ) \label{CTR}
\end{eqnarray}
Where $a$, $b$ and $c$ are the lattice parameters along the
coordinates $x$, $y$ and $z$. $q_{x}$, $q_{y}$ and $q_{z}$ are the
reciprocal space coordinates and $\delta$ is the Dirac distribution.

This well known result tells us that the intensity is not zero in
between Bragg peaks in the direction perpendicular to the surface.
Such scattering rods are called Crystal Truncation Rods (CTR)
\cite{robinson86} and arise because the crystal is truncated.
However, in general, the surface morphology deviates from this ideal
perfectly ordered case. The introduction of disorder may lead to
significant changes of the scattered intensity, particularly in
anti-Bragg condition where the intensity is proportional of that
scattered by half a lattice plane, and to an increase of calculation
complexity. It is necessary to consider the scattered intensity by
objects for which sizes and distances from one to another are
correlated. For instance in the case of a vicinal surface, the
distance separating two nearest neighbor terraces from their
respective centers is completely determined by their widths.
Therefore it is impossible to describe the scattered intensity as
the multiplication of the square modulus of the form factor of the
terrace width (averaged over the widths distribution) by an
interference function. Several authors have addressed this problem
and some analytical results have been obtained assuming generic
distributions of terrace widths \cite{houston71, lent84, pukite85,
croset97, uimin97, wollschlager97}. We propose here a method based
on the paracrystal model \cite{hosemann, guinier55, Welberry85} and
an appropriate choice of the elementary scattering object at the
atomic scale to evaluate the scattered intensity from stepped or
kinked surfaces. Previous results of HAS and electron diffraction,
i.e. the two-level model, the N-level model, the vicinal surface and
the randomly stepped surface are extended to X-ray scattering. In
comparison with these elder results the contribution of deep layers
gives rise to a CTR prefactor. This one-dimensional approach is
further expanded to two dimensions for some specific surface
morphologies. All calculations are performed assuming the atomic
structure of matter (discrete values of the terrace size and the
kink-kink distance) which allows evaluating the scattered intensity
in the whole reciprocal space and not only around Bragg peaks. This
point is of importance since surface structure characterization is
usually performed measuring the CTR or during growth by measuring
the scattered intensity in anti-Bragg conditions. Therefore Small
and Wide-Angle scattering data may be analyzed with the same model.
The concept of CTR is extended to the truncation of a terrace by a
step (Terrace Truncation Rod) and to the truncation of a step edge
by a kink (Step Truncation Rod) thanks to the terrace-step-kink
model used for the description of the surface morphology.

This paper is organized as follow. The method is first presented for
a classical one-dimensional analysis of surfaces. Different surface
morphologies are successively treated: (i) the vicinal surface, (ii)
the two-level model, (iii) the N-level model and (iv) the randomly
stepped surface ($\infty$-level model), illustrating the choice of
the elementary scattering object as well as that of paracrystals.
The cases of faceted surfaces are next treated, demonstrating the
generality of the formalism. Follows a generalization to
two-dimensions for some specific cases: (i) the kinked vicinal
surface, (ii) the two-level vicinal surface and (iii) the
$\infty$-level vicinal surface (or step meandering on a vicinal
surface). In a last part, a method is briefly proposed to take into
account in the calculation the intensity scattered by the elastic
distortions induced by periodic steps and kinks below the surface.

%\section{General expression of the intensity}

\section{One dimensional analysis of scattering from surfaces}

\subsection{The vicinal surface}
\begin{center}
\begin{figure}
\includegraphics[width=13cm]{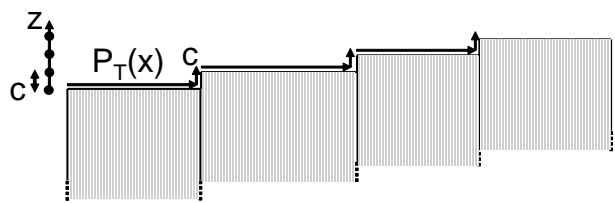}
\caption{The vicinal surface described as a stacking of terraces.
$P_{T}(x)$ is the terrace width distribution and $c$ is the step
height.} \label{stackingterraces}
\end{figure}
\end{center}

\begin{center}
\begin{figure}
\includegraphics[width=13cm]{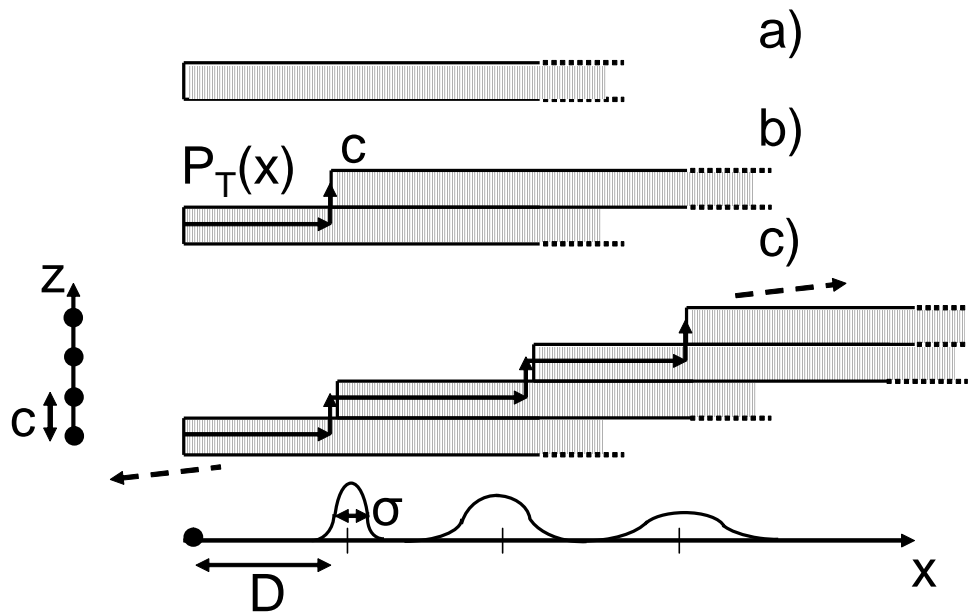}
\caption{The vicinal surface model. a) The elementary object is a
semi-infinite plane starting at the step edge. b) and c), the
vicinal surface is built step by step assuming a terrace width
distribution $P_{T}(x)$. D is the average terrace width and $\sigma$
is the FWHM. The probability laws of having a distance separating
the first, second and third neighbor steps are schematically drawn
on the x axis. Due to an increase of disorder the probability laws
broaden with the distance.} \label{vicinal}
\end{figure}
\end{center}

The usual approach for the calculation \cite{pflanz92,croset98} of
the X-ray diffraction by a vicinal surface with constant step height
is based on a description of the surface as a stacking of terraces
along the x-axis (Fig \ref{stackingterraces}). Then the scattered
intensity is averaged over all possible configurations of the step
positions. Despite its simplicity from a formal point of view, it is
necessary to take into account the correlations between the terrace
widths and the distances that separate them. However a precise
analysis of vicinal surfaces shows that an elementary object without
any size distribution can be extracted from the surface morphology.
This object consists of a semi-infinite plane parallel to the
terraces (along the x-axis), starting at the step edge (Fig
\ref{vicinal}) and penetrating into the bulk. Concerning the terrace
width distribution and the average over all the configurations, the
use of the auto-correlation function of the step positions in the
framework of the paracrystal model is straightforward. Therefore
this method allows separating in a simple way the role played by the
step structure from that played by the terrace width distribution.
In the framework of this model, the scattered intensity is the
product of the square modulus of the form factor of the elementary
object $\widetilde{F}_{vicinal}(q_{x},q_{z})$ by an interference
function $S(q_{x},q_{z})$.

\begin{equation}
I(q_{x},q_{z})=\mid \widetilde{F}_{vicinal}(q_{x},q_{z})\mid ^{2}
\times S(q_{x},q_{z})
\end{equation}

The form factor of the elementary object is a semi-infinite sum,
which yields the usual Crystal Truncation Rod behavior parallel to
the terrace width:
\begin{eqnarray}
\widetilde{F}_{vicinal}(q_{x},q_{z}) &=& \widetilde{F}_{cell}(q_{x},q_{z})\sum_{n=0}^{\infty}e^{inq_{x}a-n\mu a} \nonumber \\
&=& \widetilde{F}_{cell}(q_{x},q_{z})\frac{e^{-iq_{x}a/2}}{-2i\sin
(q_{x} a/2)}
\end{eqnarray}
where $\widetilde{F}_{cell}(q_{x},q_{z})$ is the structure factor of
the unit cell, $a$ is the lattice parameter along the x axis and
$\mu$ is the projected absorption coefficient along the x axis. One
must remember that the above formula is not correct near Bragg peaks
of perfect crystals due to multiple diffraction effects. The second
part of the calculation concerns the interference function, i.e. the
Fourier transform of the auto-correlation function of all the step
positions. The autocorrelation function, $g(x,z)$, is obtained as an
infinite summation over $n$ of the probabilities $P_{n}(x,z)$ of
having a $n^{th}$ neighbor step ($A_{n}$) at a distance (x,z) from a
step $A_{0}$ arbitrary located at the origin
$(P_{0}(x,z)=\delta(x)\times\delta (z))$.

\begin{equation}
g(x,z)=\sum_{n= -\infty}^{\infty} P_{n}(x,z)
\end{equation}

To calculate each $P_{n}(x,z)$ term, we assume that the distance
between nearest neighbor steps is described by a probability law,
the terrace widths distribution $P_{T}(x)$, and that there is no
correlation between the sizes of neighboring terraces. Within this
assumption, $P_{n}(x,z)$ can be obtained step by step building the
vicinal surface. The probability for the first neighbor step $A_{1}$
to be located at the (x,z) coordinate is
$P_{1}(x,z)=P_{T}(x)\times\delta (z-c)$, where c is the step height.
The probability of having a second neighbor step $A_{2}$ at (x',z')
results from the multiplication of the probability of having a first
neighbor step $A_{1}$ located at (x,z) by the probability of having
$A_{2}$ and $A_{1}$ separated by the vector (x'-x,z'-z) integrated
over all intermediate $A_{1}$ positions. This probability is simply
the convolution product of two probability laws : $P_{1}(x,z)\otimes
P_{1}(x,z)=P_{1}(x,z)^{\otimes 2}$. This result can be generalized
for the $n^{th}$ neighbor step :
\begin{equation}
P_{n}(x,z)=[P_{1}(x,z)]^{\otimes n}=[P_{T}(x)\times\delta
(z-c)]^{\otimes n}
\end{equation}
After adding the contribution of steps located on the negative x
axis, the interference function is obtained by Fourier transform of
the auto-correlation function of the step positions :

\begin{eqnarray}
S(q_{x},q_{z})&=&FT\left \{ g(x,z)\right \} \nonumber \\
&=&FT\left \{\delta(x)\times\delta (z)+\sum_{n= 1}^{\infty}\left
([P_{T}(x)\times\delta (z-c)]^{\otimes^{n}}+[P_{T}(-x)\times\delta
(z+c)]^{\otimes^{n}}\right ) \right \} \nonumber \\
%S_(q_{x},q_{z})&=&TF\{ \delta(x)\otimes\delta (z)+ \nonumber \\
%&+&P_{T}(x)\otimes\delta (z-h) + P_{T}(x) \otimes\delta (z-h)\otimes P_{T}(x)\otimes\delta (z-h)+ \nonumber \\
%&+&P_{T}(-x)\otimes\delta (z+h) + P_{T}(-x) \otimes\delta (z+h)\otimes P_{T}(-x)\otimes\delta (z+h)... \} \nonumber \\
S(q_{x},q_{z})&=& 1+ (\widetilde{P}_{T}(q_{x})e^{iq_{z}c} +
(\widetilde{P}_{T}(q_{x})e^{iq_{z}c})^2 ...+c.c.)
\end{eqnarray}

where
$\widetilde{P}_{T}(q_{x})=\int_{-\infty}^{+\infty}P_{T}(x)e^{iq_{x}x}dx$
is the characteristic function of the terrace widths distribution
and c.c. is the complex conjugate part resulting from the
contribution of steps located on the negative x axis. The geometric
summation converges as $\mid \widetilde{P}_{T}(q_{x}) \mid <1$ and
one gets :

\begin{eqnarray}
S(q_{x},q_{z})&=&Re\left \{
\frac{1+e^{iq_{z}c}\widetilde{P}_{T}(q_{x})
}{1-e^{iq_{z}c}\widetilde{P}_{T}(q_{x})} \right \}
\label{paracrystal}
\end{eqnarray}
Where $Re$ stands for the real part. Therefore the scattered
intensity is :

\begin{eqnarray}
I(q_{x},q_{z})&=&\left [\frac{\mid
\widetilde{F}_{cell}(q_{x},q_{z})\mid }{2\sin
{(q_{x} a/2)}}\right ]^2 Re\left \{\frac{1+e^{iq_{z}c}\widetilde{P}_{T}(q_{x})}{1-e^{iq_{z}c}\widetilde{P}_{T}(q_{x})}\right \} \nonumber \\
&=&\left [\frac{\mid \widetilde{F}_{cell}(q_{x},q_{z})\mid}{(2\sin
{(q_{x} a/2)}2\sin {(q_{z} c/2)})}\right ]^2 Re\left
\{2\frac{(1-e^{iq_{z}c})(1-\widetilde{P}_{T}(q_{x}))}{1-e^{iq_{z}c}\widetilde{P}_{T}(q_{x})}\right
\} \label{eqvicinal1}
\end{eqnarray}

This expression has been obtained by Pukite et al. (see Eq. 18-19 of
Ref \cite{pukite85}) or for instance by Rao et al. in the special
case of the geometric distribution (see Eq. 16 of Ref \cite{rao91}).
Now an additional prefactor $\frac{1}{2\sin(q_{z}c/2)^2}$ (the CTR
term) corresponding to the contribution of deep layers has to be
included. Moreover the usual prefactor $\frac{1}{q_{z}^2}$ is
replaced by $\frac{1}{\sin(q_{z}c/2)^2}$ to account for the
discontinuous structure of the lattice. Equation \ref {eqvicinal1}
highlights the symmetric roles played by steps and terraces as
$e^{iq_{z}c}$ and $\widetilde{P}_{T}(q_{x})$ are respectively the
characteristic functions of the step height and the terrace widths
distribution. As shown in Appendix A, this result can be generalized
for an arbitrary distribution of step heights $P_{S}(z)$.

\begin{equation}
I(q_{x},q_{z})=\left [\frac{\mid
\widetilde{F}_{cell}(q_{x},q_{z})\mid}{(2\sin {(q_{x} a/2)}2\sin
{(q_{z} c/2)})}\right ]^2 Re\left
\{2\frac{(1-\widetilde{P}_{S}(q_{z}))(1-\widetilde{P}_{T}(q_{x}))}{1-\widetilde{P}_{S}(q_{z})\widetilde{P}_{T}(q_{x})}\right
\} \label{eqvicinal2}
\end{equation}

where $\widetilde{P}_{S}(q_{z})$ is the characteristic function of
the step height distribution. The pre-factors $\frac{1}{2\sin
{(q_{z} c/2)}}$ and $\frac{1}{2\sin {(q_{x} a/2)}}$ can be assigned
respectively to the ideal truncation of an infinite crystal by a
flat surface, the so called CTR (see Eq. \ref{CTR}), and to the
truncation of an infinitely large terrace by a step. In analogy to
the CTR terminology, the $\frac{1}{2\sin {(q_{x} a/2)}}$ term can be
called Terrace Truncation Rod (TTR). Therefore, there is no
intensity extinction at $q_{x}a=(2n+1)\pi$, i.e. in anti-bragg
condition perpendicular to the step edges. Going further, a high
sensitivity to the step edges roughness is expected by the
combination of anti-bragg conditions both in $q_{x}$ and $q_{z}$
directions. This point will be discussed in more details within the
two-level vicinal surface model. We have to add however that the
scattered intensity may only be measurable for large enough miscut
angles as the integrated intensity (i.e. in case of long range
order) in these experimental conditions is proportional to the
number of atoms located at step edges.

\begin{center}
\begin{figure}
\includegraphics[width=13cm]{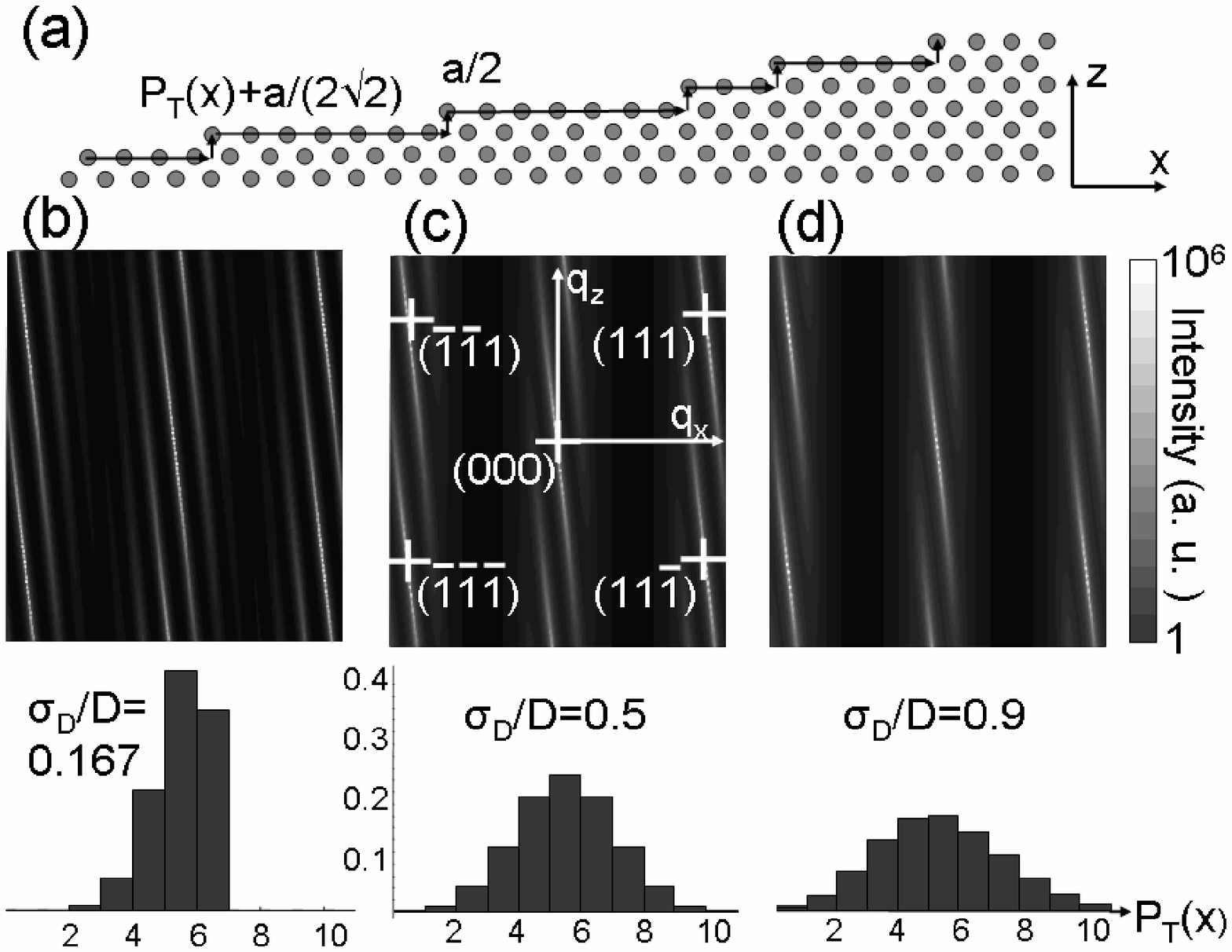}
\caption{(a) Side view of a $(115)$ vicinal surface of a fcc crystal
and scattered intensity (logarithmic scale) in the ($q_{x}0q_{z}$)
plane for different terrace width distributions. From (b) to (d) the
terrace width distribution gets larger and larger, making the
scattering rods less and less narrow. $a$ is the lattice parameter,
$a/2$ is the step height and $a/(2\sqrt{2})$ is the shift in $x$ of
the lattice parameters of two adjacent terraces.}
\label{fcc-vicinal}
\end{figure}
\end{center}

To highlight the result given by Eq. \ref{eqvicinal1}, let's
consider the intensity scattered by a vicinal surface $(11n)$ of a
fcc crystal (Fig \ref{fcc-vicinal}). The intensity is calculated in
the plane of the reciprocal space perpendicular to the step edge
direction ($q_{x}0q_{z}$). The atoms not located on the side view
section (at $y \neq 0$) are included into the calculation thanks to
the structure factor of the unit cell. Monoatomic steps are assumed
with different terrace widths distributions. The distributions are
chosen discrete by hypothesis to perform quantitative calculations
in the whole reciprocal space (Appendix B). Of course if the atomic
resolution is not required continuous distributions are enough. The
additional shift in the $x$ direction for adjacent terraces
($\frac{a}{2\sqrt{2}}$) is mandatory to keep the fcc lattice
structure. It is taken into account in the calculation adding a
phase shift, $e^{iq_{x}\frac{a}{2\sqrt{2}}}$, to the characteristic
function of the terrace width distribution. As the FWHM of the
terrace size distribution gets larger, the scattering rod gets
broader and second and third order satellites disappear in agreement
with the increase of disorder. Of course depending on the miscut
angle, the scattering rods are more or less tilted as they follow
the direction perpendicular to the macroscopic surface plane. This
remark is absolutely true close to a Bragg peak, however nonlinear
rods may exist in some cases at larger wavevector transfer. For
instance it will be shown at the end (Fig
\ref{2levelvicinaldiffraction}) that a non integer value of the mean
terrace size yields a S shape of the scattering rods.

\subsection{The two-level model}

\begin{center}
\begin{figure}
\includegraphics[width=13cm]{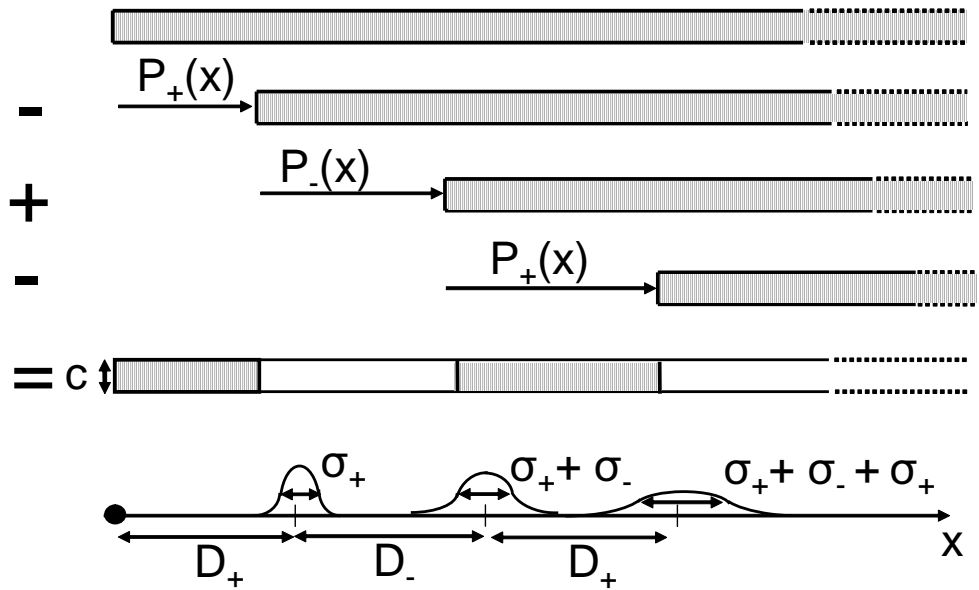}
\caption{The two-level model is built step by step assuming
alternating upward and downward steps. Terrace width distributions
are respectively $P_{+}(x)$ and $P_{-}(x)$ for the high and the low
levels. $D_{\pm}$ and $\sigma_{\pm}$ are respectively the average
terrace width and the FWHM for both distributions. The probabilities
of finding a distance $x$ separating a step up from a step down (or
up) are schematically drawn on the x-axis. } \label{2level}
\end{figure}
\end{center}

\begin{center}
\begin{figure}
\includegraphics[width=13cm]{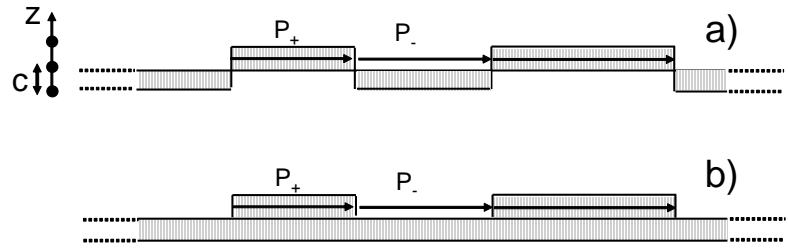}
\caption{a) Hard-Wall model (e.g. used for Helium Atom Scattering)
for which only the last atomic layer is taken into account in the
calculation of the scattered intensity. It results in an additional
($1-\cos (q_{z}c)$) term in the calculation of the scattered
intensity. b) Model for X-ray scattering for which the whole sample
volume has to be considered due to the very weak cross section. The
difference in the off-specular scattering calculation consists in
considering the atomic layer below the upper terraces.}
\label{Hard-Wall}
\end{figure}
\end{center}

The two-level model has been widely used in the literature to
describe the intensity scattered by a surface during the
homo-epitaxial growth in the sub-monolayer regime. The two level
model allows evaluating the intensity scattered by a surface with a
small roughness. The first models developed (e.g. $\beta $ model
\cite{robinson86, vlieg88}) do not provide any information on the
line shape of the diffuse scattering parallel to the surface except
a so called correlation length. Improvements have allowed analyzing
the line shape assuming a terrace widths distribution \cite{lent84,
pflanz92, croset97}. In this section, we revisit the previous
calculations. The two-level model consists in alternating upward and
downward steps (Fig \ref{2level}). As for the vicinal surface, we
consider the same elementary object, i.e. a semi-infinite sum of
unit cells parallel to the terrace plane. That way we can built step
by step a two-level model folding nodes alternatively with this
elementary object and its exact opposite (opposite electronic
density). Their form factors are opposite of each other and reads:
\begin{equation}
\widetilde{F}_{\pm}(q_{x},q_{z})=\pm
\widetilde{F}_{cell}(q_{x},q_{z})\sum_{n=0}^{\infty}e^{inq_{x}a}=\pm
\widetilde{F}_{cell}(q_{x},q_{z})\frac{e^{-iq_{x}a/2}}{-2i\sin
{(q_{x} a/2)}}
\end{equation}
Assuming that $P_{+}(x)$ and $P_{-}(x)$ are respectively the upper
and lower terrace widths distributions, the auto-correlation
function of the electronic density reads for an upward step
$F_{+}(x,z)$ as starting object and for $x>0$:
\begin{eqnarray}
g_{+}(x,z)&=& \frac{1}{2}\Big [F_{+}(x,z)\otimes F_{+}(x,z)\otimes
\delta(x) + \nonumber \\
&+& F_{+}(x,z)\otimes F_{-}(x,z)\otimes P_{+}(x) +
\nonumber \\
 &+& F_{+}(x,z)\otimes F_{+}(x,z)\otimes P_{+}(x) \otimes
P_{-}(x) + \nonumber \\
&+& F_{+}(x,z)\otimes F_{-}(x,z)\otimes P_{+}(x) \otimes
P_{-}(x)\otimes P_{+}(x)+... \Big]\nonumber \\
&=&\frac{1}{2}\Bigg [F_{+}(x,z)\otimes F_{+}(x,z)\otimes \left
(\sum_{n=0}^{\infty}\left [P_{+}(x) \otimes P_{-}(x) \right
]^{\otimes^{n}}\right )+ \nonumber \\
&+& F_{+}(x,z)\otimes F_{-}(x,z)\otimes P_{+}(x) \otimes \left
(\sum_{n=0}^{\infty}\left [P_{+}(x) \otimes P_{-}(x) \right
]^{\otimes^{n}}\right )\Bigg ]
\end{eqnarray}
The prefactor $1/2$ is used because the upward step contribution to
the auto-correlation function as starting object has to be shared
exactly with the downward step contribution. Therefore a similar
$g_{-}(x,z)$ term is added considering a downward step $F_{-}(x,z)$
as starting object. Adding also the contribution of the steps
located at $x<0$, then the scattered intensity reads :
\begin{eqnarray}
I(q_{x},q_{z})=2Re \Bigg \{\frac{1}{2}\Bigg [&\mid &
\widetilde{F}_{+}(q_{x},q_{z})\mid^2 \left (\sum_{n=0}^{\infty}\left
[\widetilde{P}_{+}(q_{x}) \widetilde{P}_{-}(q_{x}) \right
]^{^{n}}-\frac{1}{2}\right )+ \nonumber \\
&+&\mid \widetilde{F}_{-}(q_{x},q_{z})\mid^2 \left
(\sum_{n=0}^{\infty}\left [\widetilde{P}_{-}(q_{x})
\widetilde{P}_{+}(q_{x}) \right
]^{^{n}}-\frac{1}{2}\right )+ \nonumber \\
&+& \widetilde{F}^*_{+}(q_{x},q_{z}) \widetilde{F}_{-}(q_{x},q_{z})
\widetilde{P}_{+}(q_{x}) \left (\sum_{n=0}^{\infty}\left
[\widetilde{P}_{+}(q_{x}) \widetilde{P}_{-}(q_{x}) \right
]^{{n}}\right )+\nonumber \\
&+& \widetilde{F}^*_{-}(q_{x},q_{z}) \widetilde{F}_{+}(q_{x},q_{z})
\widetilde{P}_{-}(q_{x}) \left (\sum_{n=0}^{\infty}\left
[\widetilde{P}_{-}(q_{x}) \widetilde{P}_{+}(q_{x}) \right
]^{{n}}\right )\Bigg ]\Bigg \}\nonumber \\
I(q_{x},q_{z})&=&\left [\frac{\mid
\widetilde{F}_{cell}(q_{x},q_{z})\mid}{2\sin (q_{x} a/2)}\right ]^2
Re\left
\{\frac{(1-\widetilde{P}_{+}(q_{x}))(1-\widetilde{P}_{-}(q_{x}))}{1-\widetilde{P}_{+}(q_{x})\widetilde{P}_{-}(q_{x})}\right
\} \label{eq2leveldiffraction}
\end{eqnarray}
A similar result has been obtained in the case of the stacking of
layers \cite{Hermans44, hosemann}, and the exact result has already
been found for instance by Pukite et al. (see Eq. 35 of Ref
\cite{pukite85}), as well as by Croset and de Beauvais (see Eq. 20
of Ref \cite{croset97}). In these elder results, an additional
$\left [1-\cos (q_{z}c)\right ]$ dependence of the scattered
intensity was obtained which originated from the hard-wall model
hypothesis used for Helium scattering. For X-ray scattering it is
exactly canceled by the CTR prefactor,
$\frac{1}{2\sin(q_{z}c/2)^2}$, i.e. the structure factor of the
semi-infinite column perpendicular to the surface plane (Fig
\ref{Hard-Wall}). LEED patterns should be described as an
intermediate case because a penetration length of a few atomic
layers must be considered in the calculation. One can also see from
Eq. \ref{eq2leveldiffraction} that the X-ray intensity is not zero
at $q_{z}=0$. This is always the case for Helium or electron
diffraction. Qualitatively one can understand the difference that
way: for X-rays at $q_{z}=0$ , i.e. in constructive interference
conditions, one has to sum the electronic density of the atomic
planes at different $z$ coordinates deep into the bulk. As they have
different coverage (for instance in case of a multi-level surface)
each atomic plane contributes differently and the final sum does not
give rise to a homogeneous atomic plane $a$ $priori$. Therefore some
intensity is expected in $q_{z}=0$. For LEED or Helium scattering as
only the last atomic layer contributes, the sum over $z$ of the
electronic density of each atomic plane gives a perfectly
homogeneous atomic plane and therefore no off-specular intensity in
$q_{x}$.

The above calculation for X-rays is restricted to the off-specular
intensity ($q_{x}\neq 0$). The specular intensity reads :
\begin{equation} I(0,q_{z})=\mid \widetilde{F}_{cell}(0,q_{z})\mid^2\left| \frac{e^{iq_{z}c/2}}{2i\sin {(q_{z} c/2)}} +\theta
e^{iq_{z}c}\right |^2
\end{equation}
where $\theta$ is the coverage. It only depends on the upper and
lower terraces mean widths, which can be expressed in terms of the
derivative at $q=0$ of the characteristic functions of the terraces
widths distributions : $\theta
=\widetilde{P}'_{+}(0)/(\widetilde{P}'_{+}(0)+\widetilde{P}'_{-}(0))$.
Due to finite instrumental resolution and finite size effects, the
separation of specular and off-specular contributions may be
difficult to achieve. Measurements closed to anti-bragg conditions
provide a powerful tool to solve this problem as the intensity is
alternatively maximum off-specular and on the specular rod depending
on the coverage. We refer to the work of Croset and de Beauvais
\cite{croset97} for a detailed study of this case.

\subsection{The N-level model}
\begin{center}
\begin{figure}
\includegraphics[width=16cm]{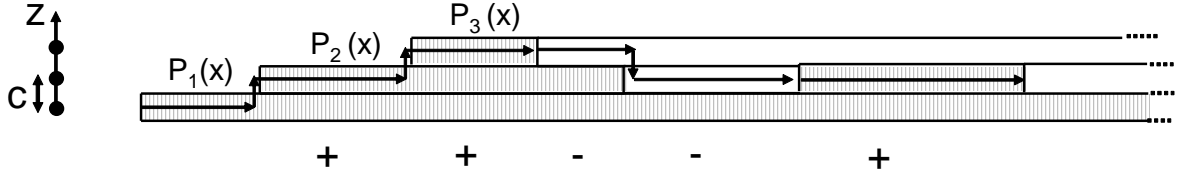}
\caption{The 3-level model is built assuming that from level 1 and
level 3, the neighboring steps must be respectively upward and
downward. From level 2, both upward and downward steps are equally
probable. The terrace width distributions are $P_{1}(x)$, $P_{2}(x)$
and $P_{3}(x)$ for each level. } \label{3level}
\end{figure}
\end{center}

The previous analysis is extended further with the N-level model.
The N-level model or multi-level model is an extension of the
two-level model for an arbitrary number of levels. Despite the
simplicity of the present approach, the calculation of the scattered
intensity for any N value remains difficult to obtain explicitly.
For the sake of simplicity, only the final result for N=3 is given
(Fig \ref{3level}):
\begin{eqnarray}
I(q_{x},q_{z})&=&\left [\frac{\mid
\widetilde{F}_{cell}(q_{x},q_{z})\mid}{2\sin
(q_{x} a/2)2\sin (q_{z} c/2)}\right ]^2 \times \nonumber \\
&\times & \Bigg[ 2 Re \left \{ \frac{(1-\cos (q_{z}c))(1-
\widetilde{P}_{2}(q_{x}))(1-\frac{\widetilde{P}_{1}(q_{x})+\widetilde{P}_{3}(q_{x})}{2})}{1-\widetilde{P}_{2}(q_{x})(\frac{\widetilde{P}_{1}(q_{x})
+ \widetilde{P}_{3}(q_{x})}{2})}\right \} + \nonumber \\
&&+ 2 Re \left \{ \frac{\frac{1}{4}(1-\cos (2q_{z}c))
\widetilde{P}_{2}(q_{x})(1- \widetilde{P}_{1}(q_{x}))(1-
\widetilde{P}_{3}(q_{x}))}{1-\widetilde{P}_{2}(q_{x})(\frac{\widetilde{P}_{1}(q_{x})
+ \widetilde{P}_{3}(q_{x})}{2})} \right \}\Bigg] \label{eq3level}
\end{eqnarray}

where $\widetilde{P}_{1}(q_{x})$, $\widetilde{P}_{2}(q_{x})$ and
$\widetilde{P}_{3}(q_{x})$ are the characteristic functions of the
terrace widths distributions of the first, second and third height
levels. They are separated by monoatomic steps of height $c$ and the
probabilities for an upward or a downward step at the intermediate
level are identical and equal to $1/2$. This result has been
obtained by Pukite et al. (see Eq. 40-44-45 of ref \cite{pukite85})
without the usual prefactor for X-rays $\frac{1}{\sin (q_{z}c/2)^2}$
for the deep layers contribution and with the prefactor
$\frac{1}{(q_{x}a)^2}$ instead of $\frac{1}{\sin (q_{x}a)^2}$ which
now fully accounts for the atomic structure of the crystal. The
analysis of Eq. ~\ref{eq3level} clearly shows two distinct
contributions respectively with pre-factors $(1-\cos (q_{z}c))$ and
$(1-\cos (2q_{z}c))$, which can be assigned to the interferences
between the waves scattered by terraces separated by respectively
one and two height levels. As for the two-level model, an additional
calculation yields the specular intensity.

\begin{equation} I(0,q_{z})=\mid \widetilde{F}_{cell}(0,q_{z})\mid^2\left| \frac{e^{iq_{z}c/2}}{2i\sin {(q_{z} c/2)}}
+\theta_{2} e^{iq_{z}c}+\theta_{3} e^{2iq_{z}c} \right|^2
\end{equation}
with
\begin{eqnarray}
\theta_{2}&=&\frac{2\widetilde{P}'_{2}(0)+\widetilde{P}'_{3}(0)}{\widetilde{P}'_{1}(0)+2\widetilde{P}'_{2}(0)+\widetilde{P}'_{3}(0)}
\qquad
\theta_{3}=\frac{\widetilde{P}'_{3}(0)}{\widetilde{P}'_{1}(0)+2\widetilde{P}'_{2}(0)+\widetilde{P}'_{3}(0)}
\end{eqnarray}

$\theta_{2}$ and $\theta_{3}$ are the coverage of the second and
third levels written in terms of the characteristic functions of the
terrace width distributions for $q_{x}=0$. For larger N values the
increase of complexity of the calculation makes the Phase Matrix
Method \cite{croset98} more appropriate.

\subsection{The randomly stepped surface : the $\infty$-level model}

\begin{center}
\begin{figure}
\includegraphics[width=16cm]{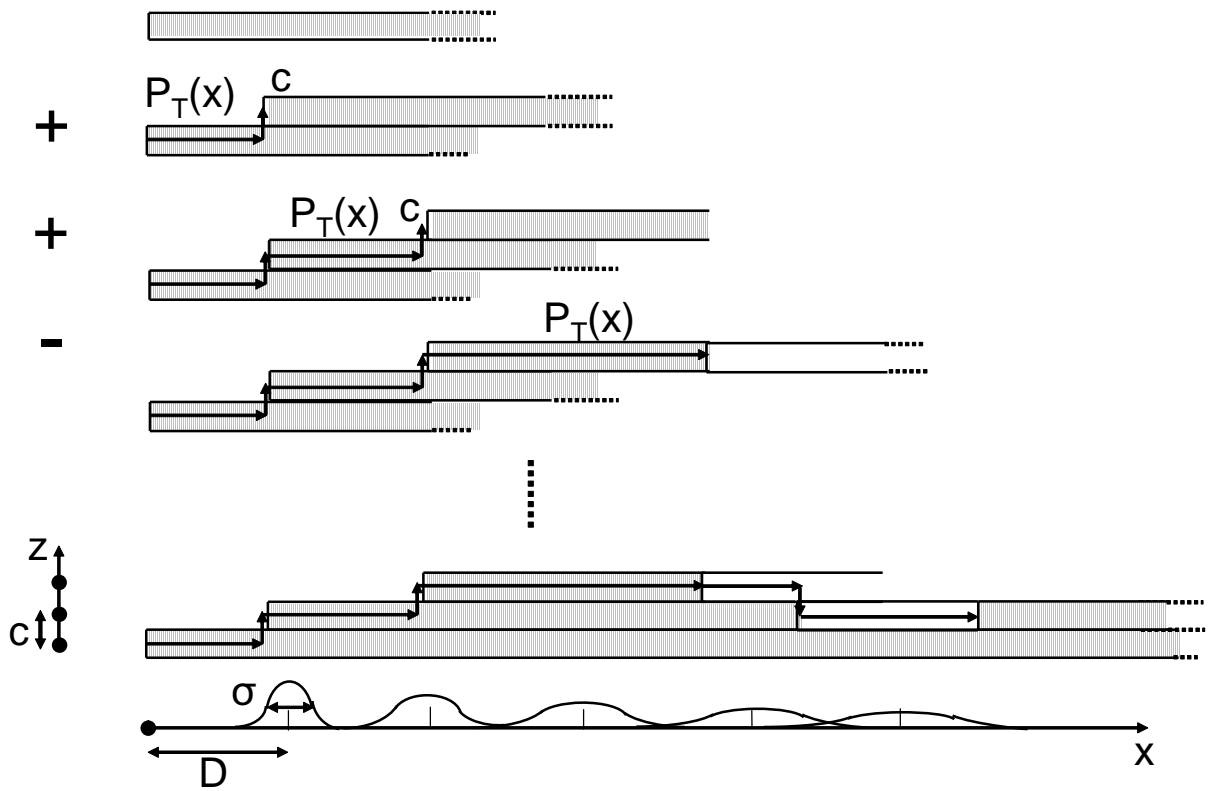}
\caption{The rough surface can be seen as an infinite level model.
The terrace width distribution is $P_{T}(x)$ $D$ is the mean
distance and $\sigma$ is the FWHM, and c is the step height. Upward
and downward steps are equally probable. As for the vicinal surface
and the two-level model, the distributions of distances separating a
$n^{th}$ neighbor step from an arbitrary step are schematically
drawn on the x-axis. The FWHM of the distributions increases with
the distance.} \label{rough}
\end{figure}
\end{center}

The above models (2 and N-level models) are limited to the analysis
of a small roughness i.e. bounded to a few atomic layers. They are
not suitable to describe the intensity scattered by a rough surface
with a divergent height correlation function. An analytic method is
proposed here (Fig \ref{rough}) to calculate the scattered intensity
assuming a random distribution of monoatomic upward and downward
steps (height c) separated by terraces having a width distribution
$P_{T}(x)$ \cite{lapujoulade81,pukite85}. Due to the divergence of
the height correlation function, the whole scattered intensity is
diffuse and no specular component has to be considered.

Similarly to the 2-level model, downward and upward steps are two
complementary objects, which in this case do not alternate regularly
but randomly with a probability $1/2$. The auto-correlation of the
electronic density calculated for an upward step as starting object
and for $x>0$ reads :

\begin{eqnarray}
%g_{+}(x,z)&=& F(x,z)\otimes F(x,z)\otimes\delta(x)\otimes\delta(z)+ \nonumber \\
%&+& F(x,z)\otimes\left [\frac{-F(x,z)\otimes\delta(z)+F(x,z)\otimes
%\delta(z-c)}{2}\right ]\otimes P_{T}(x) + \nonumber \\
%&+& F(x,z)\otimes\left [\frac{-F(x,z)\otimes\delta(z+c)+F(x,z)\otimes
%\delta(z)-F(x,z)\otimes \delta(z-c)+F(x,z)\otimes \delta(z-2c)}{4}\right ]\otimes P_{T}(x)\otimes P_{T}(x) \nonumber \\
%&+& ...\nonumber \\
g_{+}(x,z)&=& \frac{1}{2}F_{+}(x,z)\otimes F_{+}(x,z)\otimes\Big
[\delta(z)\times\delta(x)+
\left (\frac{-\delta(z)+\delta(z-c)}{2}\right )\times P_{T}(x)+ \nonumber \\
&+&\left
(\frac{-\delta(z+c)+\delta(z)-\delta(z-c)+\delta(z-2c)}{4}\right
)\times P_{T}(x)\otimes P_{T}(x)+ ...\nonumber \\
&...+& \frac{1}{2^{n+1}}\left (\sum_{k=0}^{n}\binom{n}{k}\Big (
-\delta(z+(n-2k)c)+\delta(z+(n-1-2k)c)\Big )\right )\times P_{T}(x)^{\otimes n}+...\Big ]\nonumber \\
&=& \frac{1}{2} F(x,z)\otimes F(x,z)\otimes\Big [\delta(z)\times\delta(x)+\nonumber \\
&+& \sum_{n=0}^{\infty}\frac{\sum_{k=0}^{n}\binom{n}{k}\left
(-\delta(z+(n-2k)c)+\delta(z+(n-1-2k)c) \right
)}{2^{n+1}}\times P_{T}(x)^{\otimes n}\Big ]\nonumber \\
\end{eqnarray}

where $F_{+}(x,z)$ is the shape factor of an upward step, i.e. a
semi-infinite row of atoms parallel to the terrace plane with
positive electronic density. As for the two level model the downward
steps is the opposite object, i.e. a semi-infinite row of atoms
parallel to the terrace plane with negative electronic density
$F_{-}(x,z)=-F_{+}(x,z)$. Therefore downward steps contribute in the
above expression of the autocorrelation function of the electronic
density as negative terms. The expression of the auto-correlation
function is not very intuitive but the advantage of this approach is
to get ride of the terrace width distribution. $\binom{n}{k}$ is a
binomial coefficient which accounts for the all the possible
combinations to reach a given height assuming random upward or
downward steps.

The expression of the intensity is obtained by Fourier Transform of
the auto-correlation function of the electronic density after adding
the contribution of downward steps as starting object (replacing
$\delta(z\pm nc)$ by $\delta(z\mp nc)$) in the above expression and
then including the terms for $x<0$, i. e. the complex conjugate:

\begin{eqnarray}
I(q_{x},q_{z})&=&\mid \widetilde{F}(q_{x},q_{z})\mid ^2\times  \nonumber \\
&\times & \left ( 1 +\left [i\sin
{(\frac{q_{z}c}{2})}e^{i\frac{q_{z}c}{2}}\widetilde{P}_{T}(q_{x})\sum_{n=0}^{\infty}\left
[\widetilde{P}_{T}(q_{x})\cos {(q_{z}c)}\right ]^{n} + c.c. \right ]\right )\nonumber \\
&=&\mid \widetilde{F}(q_{x},q_{z})\mid ^2\left ( 1 +\left
[\frac{i\sin
{(\frac{q_{z}c}{2})}e^{i\frac{q_{z}c}{2}}\widetilde{P}_{T}(q_{x})}{1-\widetilde{P}_{T}(q_{x})\cos
{(q_{z}c)}}+c.c. \right ]\right )
\end{eqnarray}

\begin{center}
\begin{figure}
\includegraphics[width=12cm]{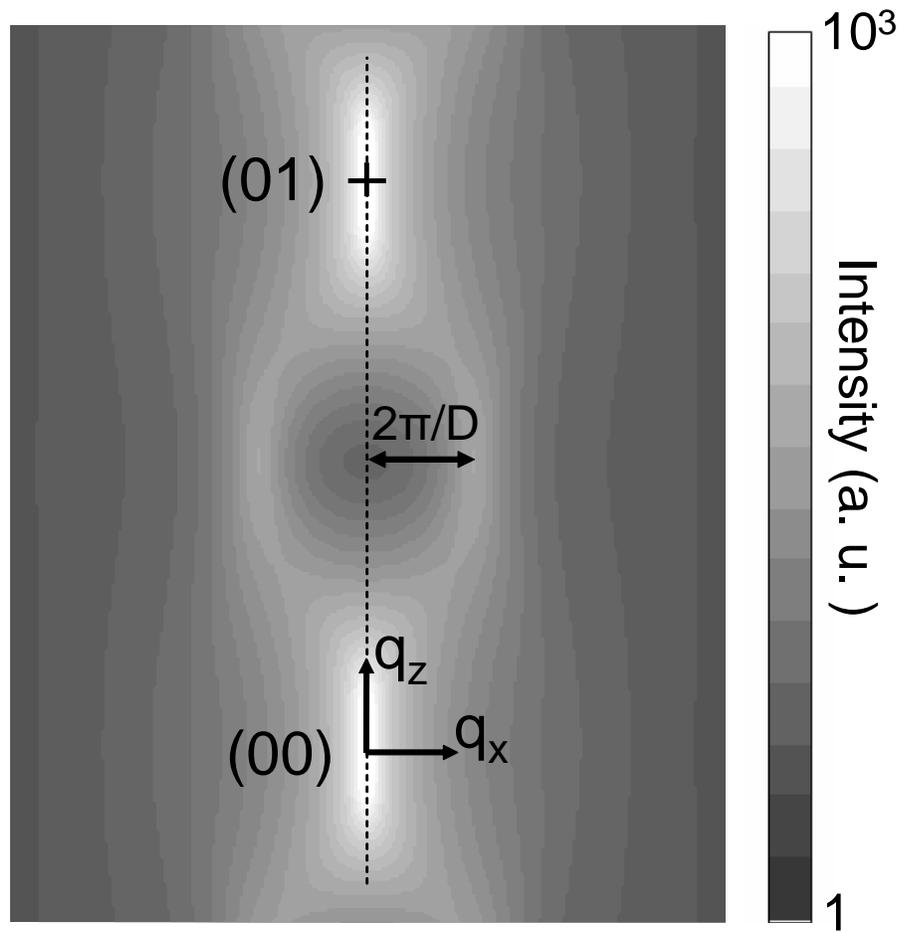}
\caption{Reciprocal space map ($q_{x},q_{z}$) of a rough surface. D
is the average distance between terraces. $\sigma / D=0.3$ (binomial
distribution, see Appendix B)} \label{rough-diffraction}
\end{figure}
\end{center}

In a more compact way:
\begin{equation}
I(q_{x},q_{z})=\left [\frac{\mid
\widetilde{F}_{cell}(q_{x},q_{z})\mid}{2\sin {(q_{x} a/2)}2\sin
{(q_{z} c/2)}}\right ]^2Re\left \{2\frac{(1-\cos
{(q_{z}c)})(1-\widetilde{P}_{T}(q_{x}))}{1-\cos{(q_{z}c)}\widetilde{P}_{T}(q_{x})}\right
\} \label{eqrough}
\end{equation}
This result is in agreement with the generalized equation of the
scattered intensity by a vicinal surface given in Eq.
\ref{eqvicinal2}. The $\cos {(q_{z}c)}$ term in Eq. \ref{eqrough} is
the characteristic function of the step heights distribution, which
is simply in this case the sum of two delta functions in $-c$ and
$+c$ weighted by $1/2$. Therefore Eq. \ref{eqvicinal2}, which is
given for a vicinal surface, can, in fact, be applied for any step
height distributions including negative height, i.e. downward steps.
This equation but without the $\frac{1}{\sin (q_{z}c/2)^2}$
prefactor has been first obtained by Lu and Lagally (see Eq. $13$ of
Ref \cite{lu82}) and then by Presicci and Lu in the special case of
a geometric terrace width distribution (see Eq. $5$ of Ref
\cite{presicci84}). Note that the scattered intensity in anti-Bragg
condition ($q_{z}c=\pi$) is the same for the randomly stepped
surface and for the two-level model at half coverage as already
mentioned by Henzler \cite{Henzler88}. This ambiguity puts in
evidence the necessity to measure the scattered intensity at
different reciprocal space positions to fully characterize the
surface roughness. For instance, in Fig \ref{rough-diffraction}, the
$q_{x}$ position of the maxima corresponding to preferential
terrace-terrace distances is a function of $q_{z}$. This result can
be compared successfully to Fig. $5$ of Pimbley and Lu
\cite{pimbley85} calculated for a truncated geometric terrace size
distribution. This can be understood through the interferences
between the waves scattered by the terrace at different heights.
This behavior does not occur for the two-level model for which the
scattered intensity (off-specular) does not depend on $q_{z}$.

To characterize a rough surface, small angle scattering techniques
are a common probe as they are preferentially sensitive to large
length scales of the surface topography. In the small momentum
transfer limit the intensity decreases as:
\begin{equation}
I(q_{x}\rightarrow 0,q_{z}\rightarrow 0)=\frac{\mid
\widetilde{F}_{cell}(q_{x},q_{z})\mid^2}{(q_{x}D)^2+\frac{(q_{z}c)^4}{4}}
\label{eqvicinal-small-angle}
\end{equation}
where D is the mean distance between nearest neighbor terraces. This
result can be compared with a more general approach based on the use
of the height correlation function \cite{villain85, sinha88,
robinson92, salanon94, legoff03}, $<(h(x)-h(0))^2>$, and in the
special case of a 1D system.

\begin{equation}
I(q_{x},q_{z})=\left [\frac{\mid
\widetilde{F}_{cell}(q_{x},q_{z})\mid}{2\sin {(q_{z} c/2)}}\right
]^2\int_{-\infty}^{+\infty}e^{-(1-\cos(q_{z}c))<(h(x)-h(0))^2>}dx
\label{eqheightcorrelation}
\end{equation}

To obtain an explicit expression the height correlation function has
to be known. In the case of the paracrystal model, the asymptotic
behavior of the height correlation function is a linear function of
the distance if the first moments of the terrace width distribution
exist (the distribution decays fast enough): $<(h(x)-h(0))^2>=
x\frac{c^2}{D}$ where c and D are respectively the mean step height
and the mean distance between adjacent steps. In this case the
calculation of the intensity from Eq. \ref{eqheightcorrelation}
reads exactly as Eq. \ref{eqvicinal-small-angle}. Of course the main
drawback of the paracrystal model is its one-dimensional character
contrary to the calculation based on the height correlation function
that holds naturally to two dimensions and is more appropriate for
instance to describe roughening \cite{villain85, legoff03}. However
for some cases as proposed in paragraph 3, the paracrystal approach
can be used to describe 2D systems. The present calculation has also
the advantage to evaluate the scattered intensity in the whole
reciprocal space (not only in the small angle regime) and for any
terrace width distributions.

\subsection{The faceted surface}

\begin{center}
\begin{figure}
\includegraphics[width=12cm]{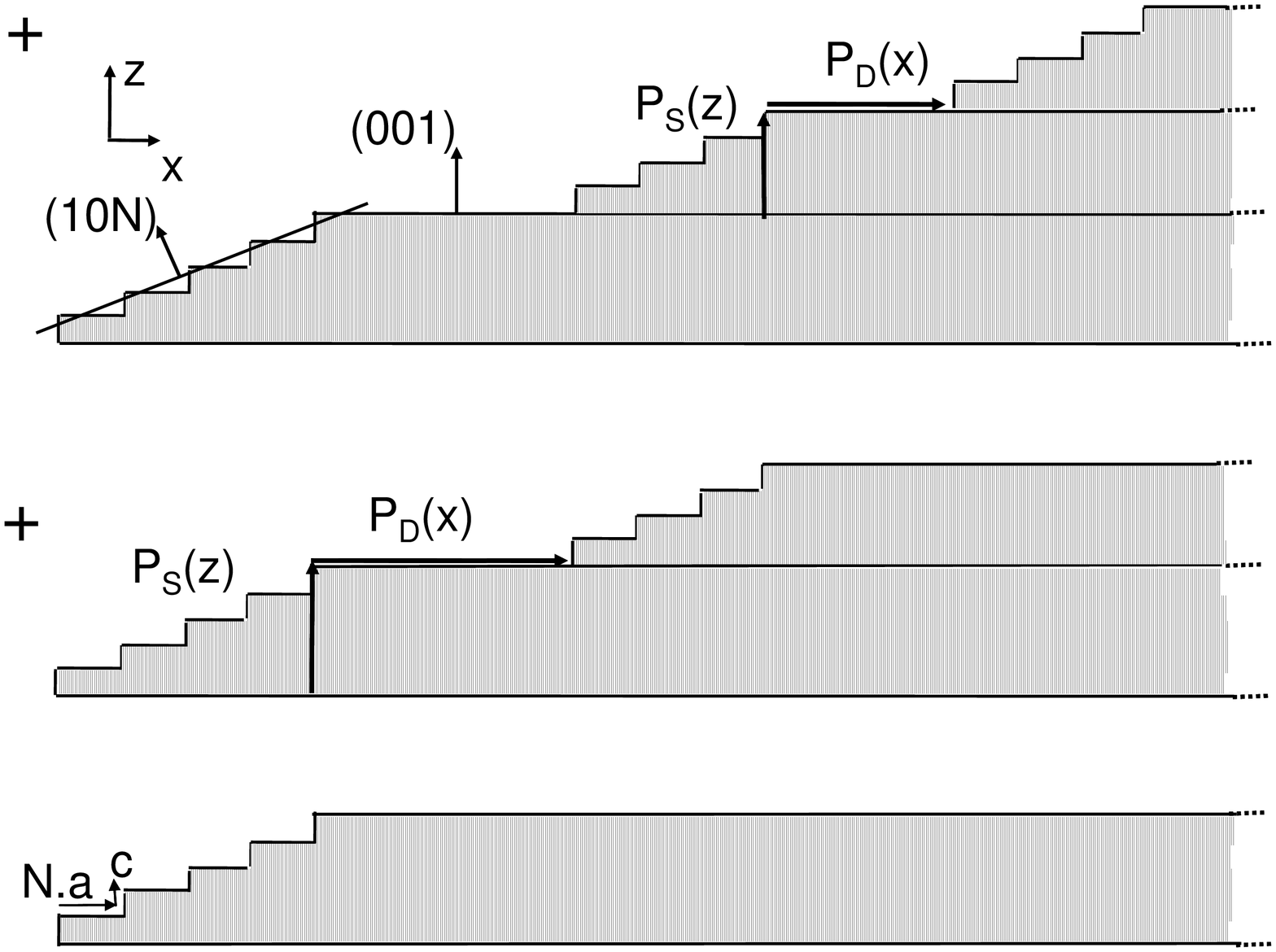}
\caption{Model of a faceted surface (small-angle faceting) showing
two types of facets, a vicinal one of (10N) type and a low index one
of (001) type. Two sets of size distributions are included: the
number of steps $P_{S}$ of the vicinal facet and the number of unit
cells $P_{D}$ of the low index facet.}
\label{facetedsurfacepetitsangles}
\end{figure}
\end{center}

\begin{center}
\begin{figure}
\includegraphics[width=12cm]{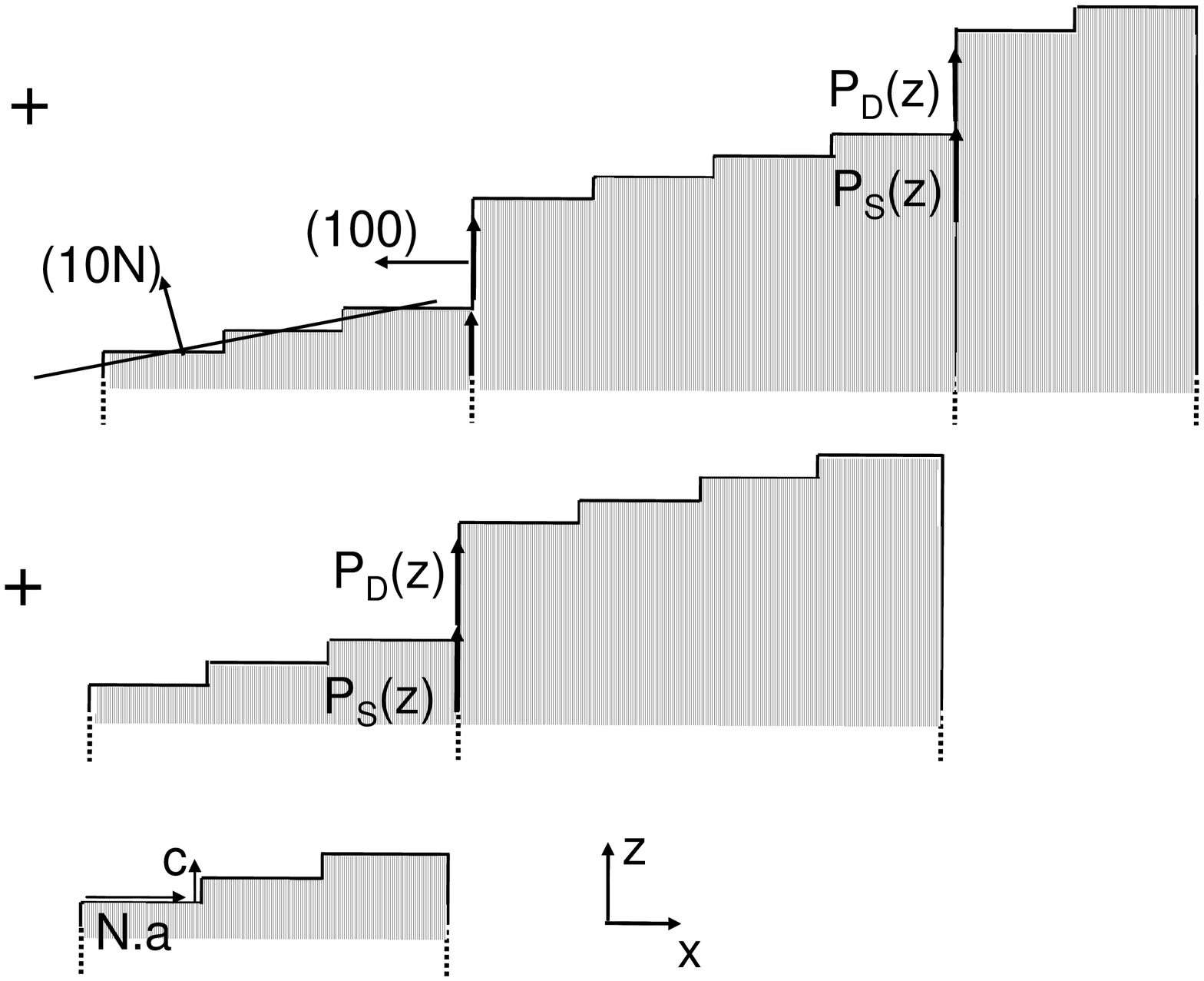}
\caption{Surface model for a large-angle faceting. Same as before
but the low index facet is a (100) type.}
\label{facetedsurfacegrandsangles}
\end{figure}
\end{center}

Surface instabilities towards faceting or step bunching have been
widely studied for instance in the case of the
electromigration-induced faceting of silicon vicinal surfaces
\cite{song95} or under adsorption of foreign species \cite{coati05}.
A model is developed here to calculate the intensity scattered by a
faceted surface exposing two types of facets with two sets of size
distribution. As an example, one of the facet type is a dense
crystallographic plane and the other one is a vicinal face. To make
simple calculations, the angle between the terrace plane of the
vicinal facet and the low index crystallographic plane of the other
facet is either 0° (Fig \ref{facetedsurfacepetitsangles}) or 90°(Fig
\ref{facetedsurfacegrandsangles}) and respectively referred to
small-angle and large-angle faceting. Similar calculations can be
performed for any other angle. The principle of the calculation is
to consider an elementary object which is a vicinal facet
semi-infinitely extended in the x-direction (resp. z-direction) for
the small-angle (resp. large-angle) faceting. For a vicinal (10N)
facet with S steps, in the small-angle faceting case, the form
factor reads:

\begin{eqnarray}
\widetilde{F}_{facet}(q_{x},q_{z}) &=& \widetilde{F}_{cell}(q_{x},q_{z})\left [ \sum_{n=0}^{\infty}e^{inq_{x}a}\right ]\left [\sum_{m= 0}^{S-1}e^{im(q_{x}Na+q_{z}c)}\right ] \nonumber \\
&=& \widetilde{F}_{cell}(q_{x},q_{z})\left
[\frac{e^{-iq_{x}a/2}}{-2i\sin (q_{x} a/2)}\right ]\left
[e^{-i\frac{(q_{x}Na+q_{z}c)(S-1)}{2}}\frac{\sin (\frac{(q_{x}
Na+q_{z}c)S}{2})}{\sin
(\frac{(q_{x}Na+q_{z}c)}{2})}\right ]\nonumber \\
\end{eqnarray}
And the interference function is (See Eq. \ref{paracrystal}):

\begin{eqnarray}
S(q_{x},q_{z})&=&Re\left \{
\frac{1+e^{i(q_{x}Na+q_{z}c)S}\widetilde{P}_{D}(q_{x})
}{1-e^{i(q_{x}Na+q_{z}c)S}\widetilde{P}_{D}(q_{x})} \right \}
\end{eqnarray}

Therefore the scattered intensity reads :

\begin{eqnarray}
I(q_{x},q_{z})&=&\left [\frac{\mid
\widetilde{F}_{cell}(q_{x},q_{z})\mid}{(2\sin {(q_{x} a/2)}2\sin
(\frac{(q_{x}Na+q_{z}c)}{2}))}\right ]^2 Re\left
\{2\frac{(1-e^{i(q_{x}Na+q_{z}c)S})(1-\widetilde{P}_{D}(q_{x}))}{1-e^{i(q_{x}Na+q_{z}c)S}\widetilde{P}_{D}(q_{x})}\right
\}\nonumber \\ \label{facetedsmallangle1}
\end{eqnarray}

The previous result can be easily checked assuming that $S=1$ and
$N=0$ which gives exactly the intensity scattered by the vicinal
surface (Eq. \ref{eqvicinal1}). A more general expression of Eq.
\ref{facetedsmallangle1} can be obtained putting instead of
$e^{i(q_{x}Na+q_{z}c)S}$, the characteristic function of the
distribution of the number of steps, $\widetilde{P}_{S}$, per
vicinal facet. It results in the following expression of the
intensity :

\begin{equation}
I(q_{x},q_{z})=\left [\frac{\mid
\widetilde{F}_{cell}(q_{x},q_{z})\mid}{(2\sin {(q_{x} a/2)}2\sin
(\frac{(q_{x}Na+q_{z}c)}{2}))}\right ]^2 Re\left
\{2\frac{(1-\widetilde{P}_{S}(q_{x}Na+q_{z}c))(1-\widetilde{P}_{D}(q_{x}))}{1-\widetilde{P}_{S}(q_{x}Na+q_{z}c)\widetilde{P}_{D}(q_{x})}\right
\} \label{facetedsmallangle2}
\end{equation}

For the large-angle faceting a very similar expression is obtained :

\begin{equation}
I(q_{x},q_{z})=\left [\frac{\mid
\widetilde{F}_{cell}(q_{x},q_{z})\mid2\sin {(q_{x} Na/2)}}{(2\sin
{(q_{x} a/2)}2\sin {(q_{z} c/2)}2\sin
(\frac{(q_{x}Na+q_{z}c)}{2}))}\right ]^2 Re\left
\{2\frac{(1-\widetilde{P}_{S}(q_{x}Na+q_{z}c))(1-\widetilde{P}_{D}(q_{z}))}{1-\widetilde{P}_{S}(q_{x}Na+q_{z}c)\widetilde{P}_{D}(q_{z})}\right
\} \label{facetedlargeangle}
\end{equation}

\begin{center}
\begin{figure}
\includegraphics[width=12cm]{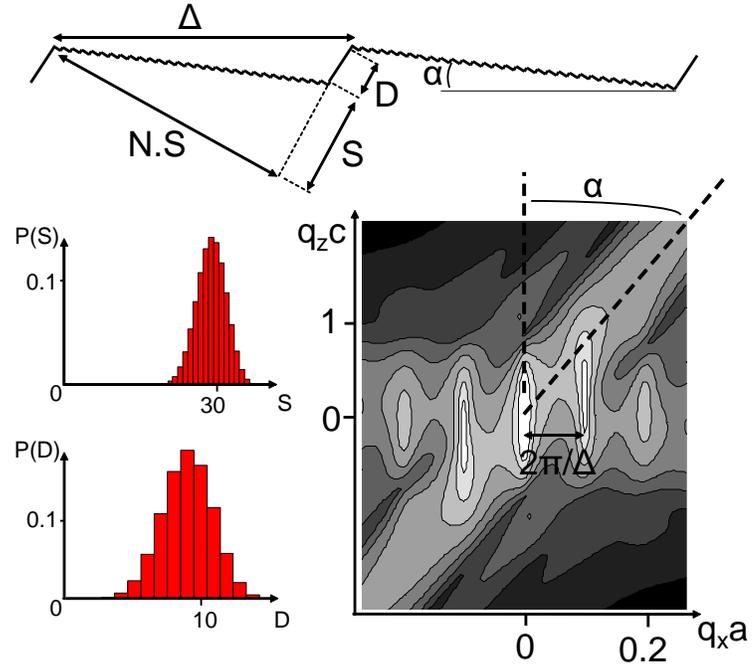}
\caption{Scattered intensity map ($q_{x},q_{z}$) in logarithmic
scale (a decade of intensity separates each color change) of a
faceted surface (large-angle faceting type). $q_{x}$ is parallel to
the macroscopic surface plane and $q_{z}$ is perpendicular. The
vicinal facet is of type (102). A schematic drawing of the surface
morphology and the size distributions of $P(S)$ and $P(D)$ are
given. $\Delta$ is the average periodicity of the faceting and
$\alpha$ is the angle between the macroscopic surface plane and the
most extended facet (vicinal facet).}
\label{facetedsurfacediffraction}
\end{figure}
\end{center}

In Fig \ref{facetedsurfacediffraction}, a simulation of the
scattered intensity close to a Bragg peak shows the main features of
the diffuse scattering by a faceted surface. The regularity of the
faceting gives rise to satellite peaks nearby the Bragg peak whereas
at larger wavevector transfer, scattering rods perpendicular to
facets are clearly visible.

\section{Towards two-dimensional analysis of scattering from surfaces}

The simplicity of the previous calculations arises because of the
one-dimensional character of the surface morphology. However, this
is also its limitation since in general surface topography should be
described in two dimensions. The appropriate generalization of this
method to two dimensions is the Markov Random Field theory as
pointed out by Lent and Cohen \cite{lent84}. Without answering this
question in the general case, a method is presented below to
calculate the intensity scattered by 2D surfaces in few important
cases : the kinked vicinal surface, the two-level vicinal surface
and the step meandering on a vicinal surface.

\subsection{The kinked vicinal surface}

\begin{center}
\begin{figure}
\includegraphics[width=13cm]{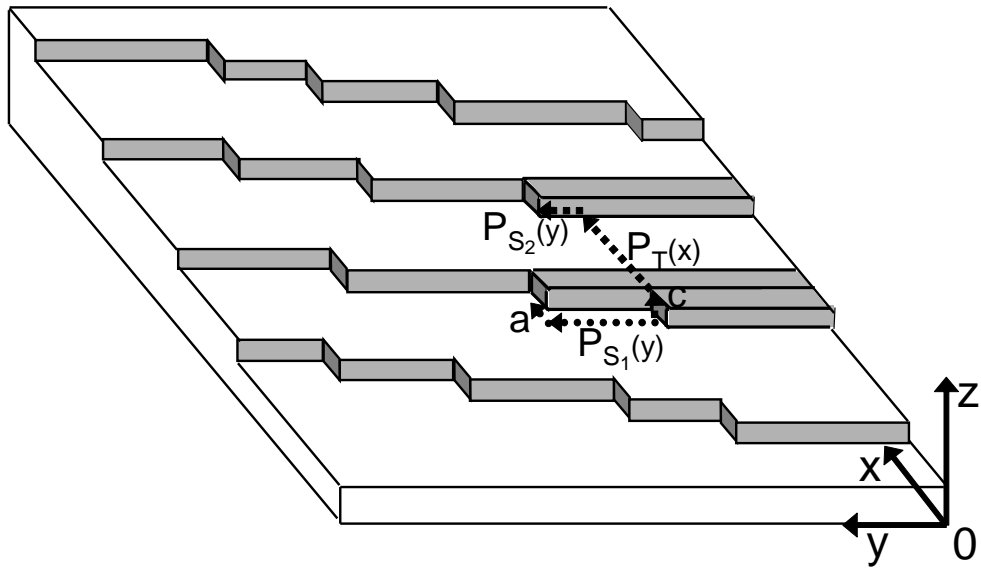}
\caption{The kinked vicinal surface model is built assuming that the
elementary object is a semi-infinite row starting at the kink edge.
Then the step edge is built from the kink-kink distance
distribution, $P_{S_{1}}(y)$, and the terrace width is controlled by
$P_{T}(x)$. $P_{S_{2}}(y)$ is an additional distribution to control
the correlations of kinks positions of neighboring steps. a and c
are respectively the kink size and the step height. }
\label{kinkedvicinal}
\end{figure}
\end{center}

Analogously to the model of the vicinal surface the first step of
the method is to find an elementary object that repeats regularly on
the kinked vicinal surface. This object is a semi-infinite row
parallel to the step (along the y-axis), starting at the kink (Fig
\ref{kinkedvicinal}). For instance, in the case of a monoatomic
kink, the form factor of the elementary object is :

\begin{eqnarray}
F_{\text{kinked vicinal}} &=& \widetilde{F}_{cell}(q_{x},q_{y},q_{z})\sum_{n=-\infty}^{0}e^{inq_{y}b + n\mu b} \nonumber \\
&=&
\frac{\widetilde{F}_{cell}(q_{x},q_{y},q_{z})e^{iq_{y}b/2}}{2i\sin
(q_{y} b/2)}
\end{eqnarray}

In order to build the whole kinked vicinal surface, an infinite
assembly of these objects are arranged along two main directions :
the step edges (by definition) and perpendicularly to the steps to
take into account any spatial correlations between kinks of
neighboring steps. Both directions are decoupled by hypothesis and
the kinks positions are given by the ideal 2D-paracrystal model
\cite{hosemann, eads01}, assuming distance distributions between
nearest neighbor kinks. However kinks (and steps) interact with each
other via long range interactions (elastic, electrostatic) that
yield a well defined separation between neighboring kinks. To solve
this problem it would be necessary to take into account correlations
with far neighbors which would make the calculation much more
difficult. Our purpose is to give a simple and analytical
calculation of the intensity scattered by a kinked vicinal surface
assuming generic distributions of distances between nearest
neighbors kinks. Of course a relevant choice of the distribution
laws may allow partially taking into account the kink interaction.

In the direction parallel to the step edges the interference
function is :

\begin{equation}
S_{\parallel}(q_{x},q_{y})=Re\left
\{\frac{1+e^{iq_{x}a}\widetilde{P}_{{S_{1}}}(q_{y})}{1-e^{iq_{x}a}\widetilde{P}_{{S_{1}}}(q_{y})}\right
\}
\end{equation}

where $\widetilde{P}_{{S_{1}}}(q_{y})$ is the characteristic
function of the distance separating two nearest neighbor kinks along
the step edge and ${a}$ is the kink width (monoatomic in this case).
This calculation is identical to the one for the vicinal surface
model (see Eq. \ref{paracrystal}).

Similarly in the direction perpendicular to the steps, the
interference function is given by:
\begin{equation}
S_{\perp }(q_{x},q_{y},q_{z})=Re\left
\{\frac{1+e^{iq_{z}c}\widetilde{P}_{T}(q_{x})\widetilde{P}_{{S_{2}}}(q_{y})}{1-e^{iq_{z}c}\widetilde{P}_{T}(q_{x})\widetilde{P}_{{S_{2}}}(q_{y})}\right
\}
\end{equation}

where $\widetilde{P}_{T}(q_{x})$ and
$\widetilde{P}_{{S_{2}}}(q_{y})$ are respectively the characteristic
functions of the terrace widths distribution and kink positions
along the upper (lower) step edges. $\widetilde{P}_{{S_{2}}}(q_{y})$
denotes the variations of the kink positions along adjacent step
edges.

The expression of the intensity reads :

\begin{eqnarray}
I(q_{x},q_{y},q_{z})&=& \mid \widetilde{F}_{\textrm{kinked
vicinal}}\mid^{2}\times
S_{\parallel}(q_{x},q_{y})\times S_{\perp }(q_{x},q_{y},q_{z}) \nonumber \\
&=&\left [\frac{\mid \widetilde{F}_{cell}(q_{x},q_{y},q_{z})\mid
}{2\sin (q_{y}
b/2)}\right ]^2 \times \nonumber \\
&\times &Re\left
\{\frac{1+e^{iq_{x}a}\widetilde{P}_{S_{1}}(q_{y})}{1-e^{iq_{x}a}\widetilde{P}_{S_{1}}(q_{y})}\right
\} \times \nonumber \\
&\times &Re\left
\{\frac{1+e^{iq_{z}c}\widetilde{P}_{T}(q_{x})\widetilde{P}_{S_{2}}(q_{y})}{1-e^{iq_{z}c}\widetilde{P}_{T}(q_{x})\widetilde{P}_{S_{2}}(q_{y})}\right
\} \nonumber \\
&=&\left [\frac{\mid \widetilde{F}_{cell}(q_{x},q_{y},q_{z})\mid
}{2\sin (q_{x} a/2)2\sin (q_{y} b/2)2\sin (q_{z}
c/2)}\right ]^2 \times \nonumber \\
&\times &Re\left
\{2\frac{(1-e^{iq_{x}a})(1-\widetilde{P}_{S_{1}}(q_{y}))}{1-e^{iq_{x}a}\widetilde{P}_{S_{1}}(q_{y})}\right
\} \times \nonumber \\
&\times &Re\left
\{2\frac{(1-e^{iq_{z}c})(1-\widetilde{P}_{T}(q_{x})\widetilde{P}_{S_{2}}(q_{y}))}{1-e^{iq_{z}c}\widetilde{P}_{T}(q_{x})\widetilde{P}_{S_{2}}(q_{y})}\right
\} \label{eqdouble-vicinal}
\end{eqnarray}

\begin{center}
\begin{figure}
\includegraphics[width=16cm]{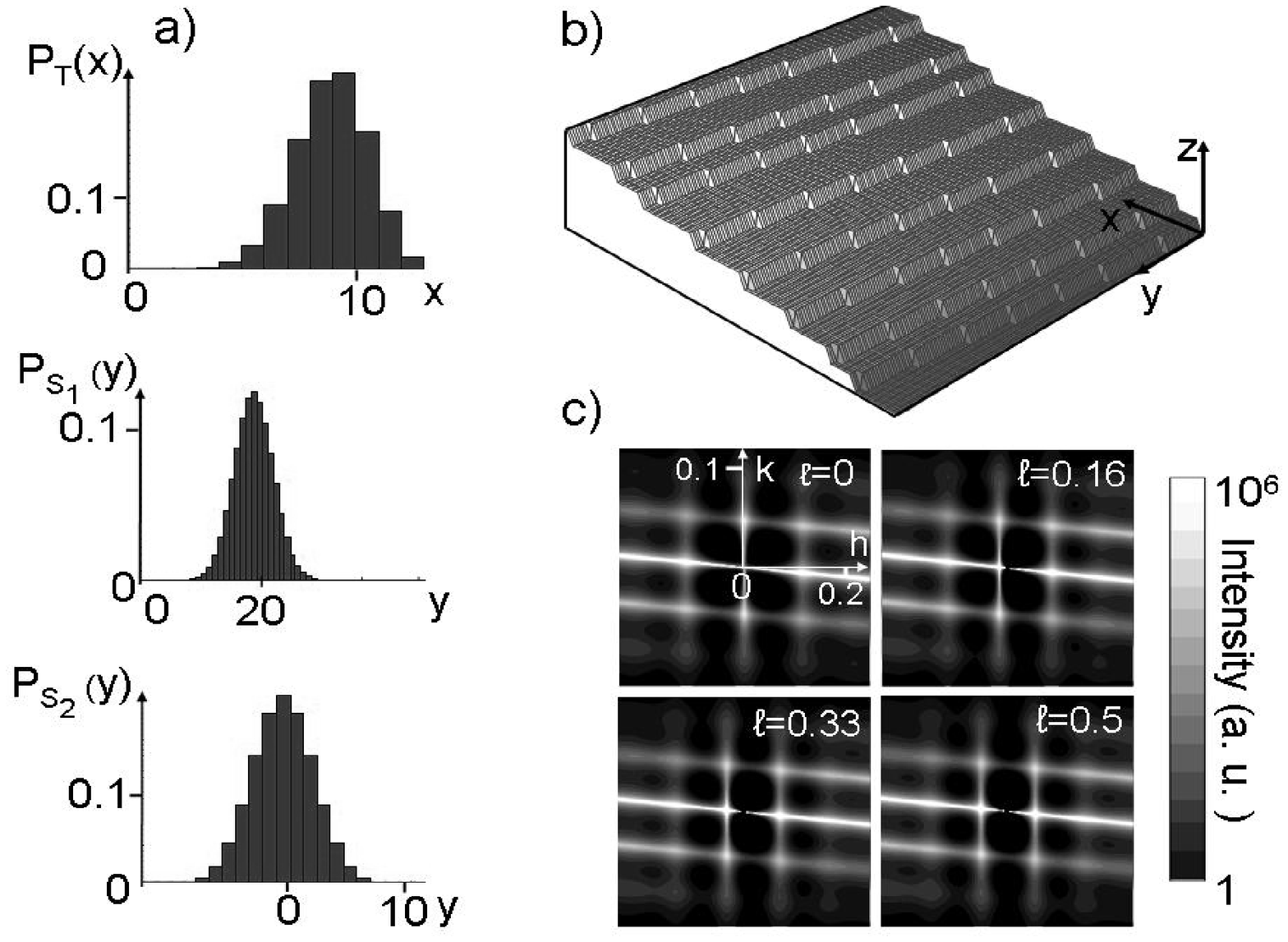}
\caption{a) Step distance distribution $P_{T}(x)$, kink-kink
distance distribution along the step edge $P_{S_{1}}(y)$, and
kink-kink distance distribution along nearest neighbor steps
$P_{S_{2}}(y)$. b) Simulation of the corresponding surface
morphology. c) Intensity map scattered by this kinked vicinal
surface at $q_{z}=0$. } \label{kinkedvicinal-test}
\end{figure}
\end{center}

The first term in Eq. \ref{eqdouble-vicinal} consists of the
multiplication of four terms : the form factor of the unit cell; the
CTR, $1/2\sin {(q_{z} c/2)}$; the Terrace Truncation Rod, $1/2\sin
{(q_{y} b/2)}$, which has already been highlighted in the case of
the vicinal surface model and a new term, $1/2\sin {(q_{x} a/2)}$,
which can be assigned to the truncation of an infinite step edge by
a kink. Following previous terminology, it can be called Step
Truncation Rod (STR). One should also notice that the scattered
intensity is calculated in the whole reciprocal space (no specular
scattering) as the whole sample volume (not only the surface) is
taken into account in the calculation. As expected the intensity
calculation shows two sets of scattering rods that are not
perpendicular to each other and which correspond to the steps and
kinks networks (Fig \ref{kinkedvicinal-test}). Due to the miscut
angle of the surface, the scattering rods shift along $q_{z}$, in
the $q_{x}$ direction (steps) as well as in $q_{y}$ direction
(kinks) in order to be perpendicular to the average surface plane.
The $q_{y}$ shift is hardly visible as the orientation of the normal
to the macroscopic surface plane depends slowly on the kinks
density. For completeness about Eq. ~\ref{eqdouble-vicinal}, a kink
distribution can be considered (e.g. for non monoatomic kinks)
replacing $e^{iq_{x}a}$ by the characteristic function of the kink
width distribution. The same can be performed for the step height
($e^{iq_{z}c}$). In addition to the topographic characterization of
a kinked vicinal surface, this model can be applied to calculate the
scattered intensity during epitaxial growth at kink edges. The form
factor of the elementary object has to be modified according to :

\begin{eqnarray}
\widetilde{F}_{\text{kinked vicinal}} &=& \widetilde{F}_{cell}(q_{x},q_{y},q_{z})\sum_{n=-\infty}^{0}e^{inq_{y}b + n\mu b} + \widetilde{F}_{a}e^{i\vec{q}\vec{r}}\nonumber \\
&=&
\frac{\widetilde{F}_{cell}(q_{x},q_{y},q_{z})e^{iq_{y}b/2}}{2i\sin
(q_{y} b/2)}+ \widetilde{F}_{a}e^{i\vec{q}\vec{r}}
\end{eqnarray}

where $\widetilde{F}_{a}$ is the structure factor of the additional
atom located at a vector $\vec{r}$ from the kink edges. Therefore
intensity variations might be observed at anti-Bragg conditions,
i.e. in the center of the reciprocal lattice nodes
($q_{x}a=q_{y}b=q_{z}c=\pi$). However this may be hardly measurable
due to the very small number of atoms involved in the scattering
process resulting from destructive interferences between the waves
scattered by all the atoms. Indeed without foreign atoms the
intensity results from only half of the atoms localized at the kinks
positions. This is similar to X-ray scattering by a flat surface in
anti-Bragg conditions where only half of the atoms of the last
surface plane give a contribution (cubic case, \cite{robinson92}).
In homoepitaxy, no intensity change is expected (the kinks cannot be
rough). However in heteroepitaxy, the intensity can vary
significantly depending on the structure factor of the foreign atom
at its position with respect to the kink site.

\subsection{The two-level vicinal surface}

\begin{center}
\begin{figure}
\includegraphics[width=13cm]{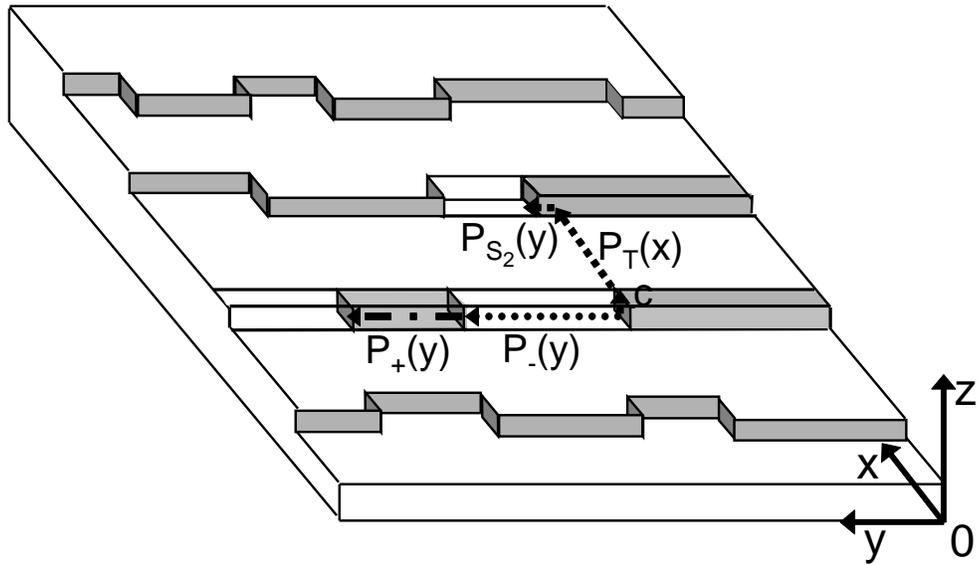}
\caption{The two-level vicinal surface model allows describing the
step-edge of a vicinal surface as a two-level model. $P_{+}(y)$ and
$P_{-}(y)$ are the width distributions of the higher and lower
levels. As for the kinked vicinal surface model, the terrace width
is controlled by $P_{T}(x)$ and the kink-kink distance along the
y-direction for neighboring steps by $P_{S_{2}}(y)$. }
\label{2levelvicinal}
\end{figure}
\end{center}

The aim of this part is to calculate the intensity scattered by a
vicinal surface which has rough step edges \cite{wollschlager98,
tegenkamp02}. For that purpose we consider that the step edges can
be described by a two-level model (Fig \ref{2levelvicinal}). Making
use of previous results for the two-level model and for the vicinal
surface the scattered intensity reads :

\begin{eqnarray}
I(q_{x},q_{y}\neq 0,q_{z})&=&\left [\frac{\mid
\widetilde{F}_{cell}(q_{x},q_{y},q_{z})\mid}{2\sin (q_{y}
b/2)}\right ]^2 \times
\nonumber \\
&\times &Re\left
\{\frac{(1-\widetilde{P}_{+}(q_{y}))(1-\widetilde{P}_{-}(q_{y}))}{1-\widetilde{P}_{+}(q_{y})\widetilde{P}_{-}(q_{y})}\right
\}\times \nonumber \\
&\times &Re\left
\{\frac{1+e^{iq_{z}c}\widetilde{P}_{T}(q_{x})\widetilde{P}_{S_{2}}(q_{y})}{1-e^{iq_{z}c}\widetilde{P}_{T}(q_{x})\widetilde{P}_{S_{2}}(q_{y})}\right
\} \label{eq2levelvicinal-nonspecular}
\end{eqnarray}
Where the first two terms at the right hand side of the equation has
already been given in the case of the two-level model (see Eq.
\ref{eq2leveldiffraction}). Indeed a two-level model for step edges
is equivalent to a one dimensional two-level model for a surface.
The third term deals with the interference function for adjacent
steps (see for instance Eq. \ref{paracrystal} of the vicinal
surface) with an additional term $\widetilde{P}_{S_{2}}(q_{y})$
which accounts for correlations between the positions of kinks at
adjacent steps.

In this case the calculation of the scattered intensity is
restricted to $q_{y}\neq 0$. Similarly to the two-level model one
has to make an additional calculation for this case :
\begin{equation} I(q_{x},0,q_{z})=\mid \widetilde{F}_{cell}(q_{x},0,q_{z})\mid^2\left| \frac{e^{-iq_{x}a/2}}{-2i\sin (q_{x} a/2)} +\theta
e^{-iq_{x}a}\right |^2 Re\left
\{\frac{1+e^{iq_{z}c}\widetilde{P}_{T}(q_{x})}{1-e^{iq_{z}c}\widetilde{P}_{T}(q_{x})}\right
\} \label{eq2levelvicinal-specular}
\end{equation}

\begin{center}
\begin{figure}
\includegraphics[width=12cm]{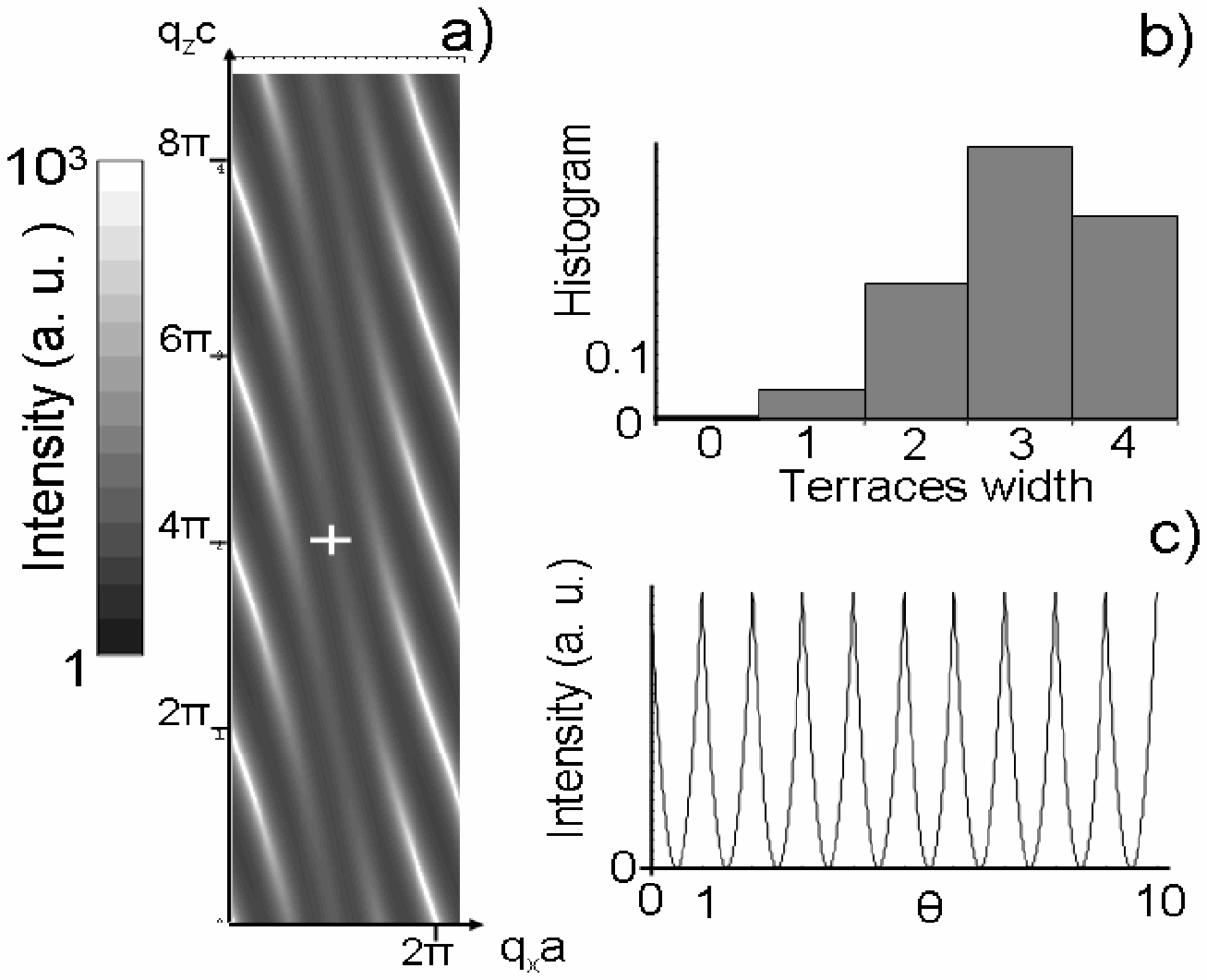}
\caption{a) Intensity map (logarithmic scale) of the two-level
vicinal surface model in the $(q_{x}0q_{z})$ plane. b) Histogram of
the terrace width used for the calculation. c) Intensity
oscillations calculated in anti-Bragg condition (white cross on a).
$\theta$ is the coverage rate of the step edge. One can notice the S
shape of the scattering rods arising from the non integer value of
the mean terrace size.} \label{2levelvicinaldiffraction}
\end{figure}
\end{center}

Where $\theta$ is the analogue of the coverage applied to a step
edge. As already mentioned in the part dedicated to the
one-dimensional analysis of vicinal surfaces, the roughness on step
edges modifies the Terrace Truncation Rod in a way similar to the
roughness on a flat surface for the CTR (Eq.
\ref{eq2levelvicinal-specular}). This remark calls for additional
comments. For instance in the case of homoepitaxial growth on a
vicinal surface, intensity oscillations may be observed in
anti-Bragg conditions if the step edges are alternatively rough and
smooth (Fig \ref{2levelvicinaldiffraction}). To maximize the
sensitivity to the roughness one has to be in anti-Bragg condition
on the scattering rod of the step network ($q_{x}a=\pi$, $q_{y}b=0$,
$q_{z}c=\pi$). Fig \ref{2levelvicinaldiffraction} reveals also that
the scattering rods have a S shape. This can be assigned to the non
integer value of the mean terrace size. Indeed, close to a Bragg
peak, the scattering rods are perpendicular to the average surface
plane. As the mean terrace size is not an integer, the scattering
rods do not point towards a Bragg peak. Therefore there is a slight
orientation change close to anti-Bragg conditions to connect two
head-to-tail scattering rods \cite{pimbley85}. In addition shoulders
in $q_{y}$ direction originating from a preferential distance
between kinks are expected in the diffuse scattering (Eq.
\ref{eq2levelvicinal-nonspecular}) as pointed out by Wollschläger et
al. \cite{wollschlager98} and revealed experimentally by Tegenkamp
et al. \cite{tegenkamp02} with SPA-LEED. This effect should be
enhanced at half coverage of the step edges (maximum roughness)
exactly in the same way as the two-level model for a surface (see
Ref \cite{croset97}). In the case of heteroepitaxial growth at step
edges on vicinal surfaces (e.g. Co/Pt(997) \cite{gambardella01})
complex interference effects are expected arising from the different
scattering power between the substrate atoms and the adatoms. This
model can be extended to the N-level vicinal surface. One has to add
that the correlation between the morphologies of neighboring step
edges may be adjusted thanks to the $\widetilde{P}_{{S_{2}}}(q_{y})$
function. It is related to the kink-kink relative displacement along
two adjacent step edges. If the kinks positions are fully
correlated, e.g. at identical $y$ position from step to step,
$\widetilde{P}_{{S_{2}}}(q_{y})=1$. If the morphologies are
independent $\widetilde{P}_{{S_{2}}}(q_{y})=\delta(q_{y})$. In this
latter case, for $q_{y}\neq 0$ the scattered intensity is the
incoherent sum of the contributions of independent step edges. For a
detailed study of this case we refer to the work of Wollschläger et
al. \cite{wollschlager98}. Any intermediate case can be considered.

\subsection{Step meandering on vicinal surfaces : the $\infty$-level vicinal surface model}

\begin{center}
\begin{figure}
\includegraphics[width=13cm]{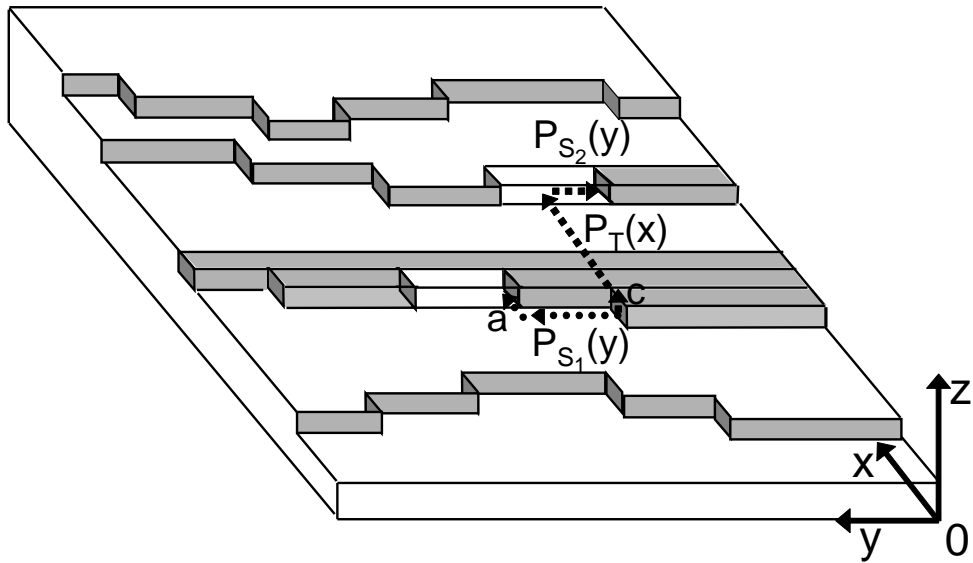}
\caption{The step meandering model on a vicinal surface is built
assuming that the step edge can be described by the infinite level
model (kink-kink distance distribution: $P_{S_{1}}(y)$). The terrace
width is controlled by $P_{T}(x)$ and the kink-kink distance along
the neighboring step edge in the y-direction by $P_{S_{2}}(y)$.}
\label{meanderingvicinal}
\end{figure}
\end{center}

The last example deals with step meandering on vicinal surfaces
\cite{villain85}. Instabilities on vicinal surfaces induced by
specific growth conditions or an external stress may lead to step
meandering \cite{muller04}. Making use of previous results on
randomly stepped surfaces and vicinal surfaces, the generalization
to step meandering on vicinal surfaces is straightforward (Fig
\ref{meanderingvicinal}). The scattered intensity for a kink size
$a$ and a step height $c$ is :

\begin{eqnarray}
I(q_{x},q_{y},q_{z})&=&\left [\frac{\mid
\widetilde{F}_{cell}(q_{x},q_{y},q_{z})\mid}{2\sin (q_{x} a/2)2\sin
(q_{y} b/2)}\right ]^2 \times
\nonumber \\
&\times &Re\left \{2\frac{(1-\cos
{(q_{x}a)})(1-\widetilde{P}_{S_{1}}(q_{y}))}{1-\cos{(q_{x}a)}\widetilde{P}_{S_{1}}(q_{y})}\right
\} \times \nonumber \\
&\times &Re\left
\{\frac{(1-e^{iq_{z}c})(1-\widetilde{P}_{T}(q_{x})\widetilde{P}_{S_{2}}(q_{y}))}{1-e^{iq_{z}c}\widetilde{P}_{T}(q_{x})\widetilde{P}_{S_{2}}(q_{y})}\right
\} \label{eqmeanderingvicinal}
\end{eqnarray}

\begin{center}
\begin{figure}
\includegraphics[width=12cm]{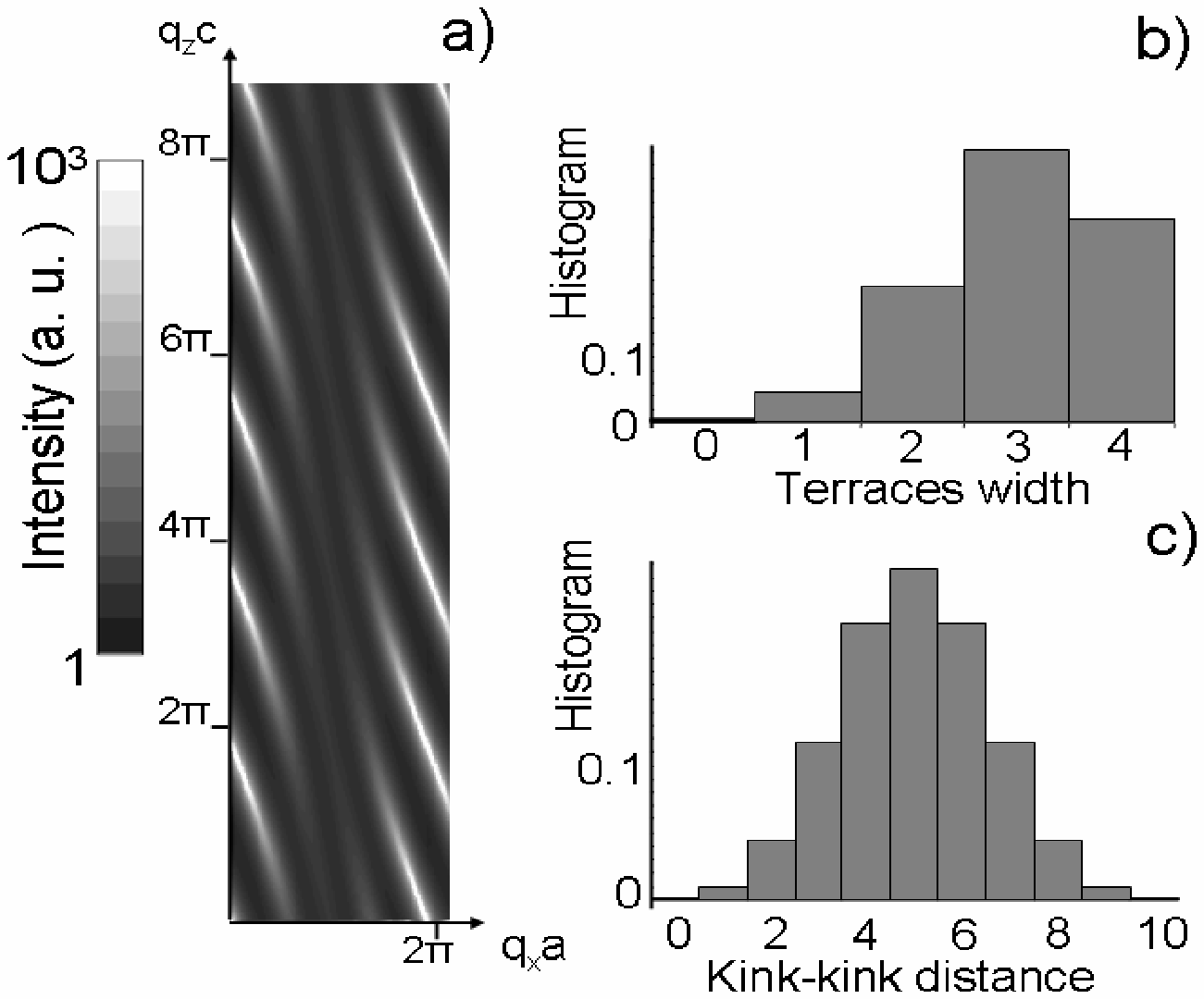}
\caption{a) Intensity map (logarithmic scale) of the step meandering
vicinal surface model in the $(q_{x}0q_{z})$ plane. b) Histogram of
the terrace width distribution. c) Histogram of the kink-kink
distance distribution : $P_{S_{1}}(y)$. $P_{S_{2}}(y)$ is identical
to $P_{S_{1}}(y)$ but centered around $y=0$. Compared to the
two-level vicinal surface, the intensity decreases much faster in
$q_{x}$ because the step edges are much more rough.}
\label{meanderingvicinaldiffraction}
\end{figure}
\end{center}

Where the first two terms on the right hand side of the equation are
directly extracted from Eq. \ref{eqrough} of the randomly stepped
surface. Indeed a step edge with random kinks is equivalent to the
one dimensional randomly stepped surface model. The last term is the
interference function of adjacent steps as seen previously for
instance in the case of the kinked vicinal surface. A simulation of
the intensity scattered by the step meandering model is shown in Fig
\ref{meanderingvicinaldiffraction}. The intensity decreases much
faster in the $q_{x}$ direction compared to the two-level vicinal
surface model ($q_{x}^{-4}$ instead of $q_{x}^{-2}$) due to very
rough step edges.

For this model one may wonder if steps may collide which is a non
physical situation due to step-step repulsion. Even if the step
distance distribution, i.e. $P_{T}(x)$, takes only positive values,
steps may collide due to the $P_{S_{2}}(y)$ probability distribution
for the kink-kink distance of adjacent steps. The effect of
$P_{S_{2}}(y)$ is to decorrelate the kink positions of two neighbor
steps. If one choose a very broad $P_{S_{2}}(y)$ probability
distribution, a step may collide with an adjacent step. However this
can be avoided by choosing a $P_{S_{2}}(y)$ probability distribution
with a maximum $\mid y \mid$ displacement much smaller than the
product of the kink-kink distance on a step edge times the number of
kinks to fill the terrace width. This condition is not very
restrictive but one can see that the limit case $P_{S_{2}}(y)=1$
does not fulfill this requirement. It leads to the diffraction by
decorrelated steps as pointed out by Wollschläger et al.
\cite{wollschlager98}, but of course uncorrelated steps with random
kinks on the step edges may collide. This remark on $P_{S_{2}}(y)$
applies also for the kinked vicinal surface model but not for the
two-level vicinal surface model as the step-step fluctuations are
confined in $\pm 1$ atomic distance and the non crossing condition
for steps is easily fulfilled.

As for the kinked vicinal surface, the calculation provides the
whole scattered intensity as their is no specular scattering. Of
course the above expression can be generalized for any kink size and
step height distributions. For instance it is straightforward to
combine the randomly stepped model with a random distribution of
kinks along the step edges just changing $e^{iq_{z}c}$ by
$\cos{(q_{z}c)}$ in Eq. \ref{eqmeanderingvicinal}.

\section{Lattice distortion}

Until now no lattice distortion have been introduced and the
crystallographic structure is the same at the surface and in the
bulk. This assumption may be the most limiting one to quantitatively
characterize surfaces with X-rays \cite{prevot03, prevot04b,
croset06} as steps and kinks are intrinsic surface defects which
stress the sample. To solve this problem, we suggest an approximate
calculation of the scattered intensity which takes into account the
strain field induced by steps and kinks assuming that the surface
morphology is not perfectly periodic. To illustrate this point,
let's consider the case of a vicinal surface. For a perfectly
periodic network of steps the strain field along only one atomic
plane parallel to the terrace and starting at a step edge allows
describing the strain field inside the whole sample volume
\cite{croset04}. This aforementioned atomic plane is nothing else
than the semi-infinite elementary object used to build the vicinal
surface model (see Fig~ \ref{vicinal}). Consequently the strain
field calculation (e.g. in the framework of the linear elasticity
theory) obtained for a perfect network of steps, may be applied to a
slightly disordered one by simply using the form factor of the
strained elementary object. Of course for highly disordered vicinal
surfaces this calculation will not be appropriate. One has to be
aware that a variation of the terrace size modifies not only the
amplitude of the strain field but also the positions of the
accidents of the displacement field which are located below the
nearest neighbor steps. Therefore much caution is necessary to use
this approximation. A similar result can be obtained for the kinked
vicinal surface. The strain field induced by kinks on an ideal
kinked vicinal surface (bi-periodic kinks network) can be obtained
through the strain field calculation along only one atomic row
parallel to a step edge and starting at a kink. Therefore an
approximate calculation of the scattered intensity by a strained
kinked vicinal surface can be obtained transferring this latter
result to the elementary object. However contrary to the vicinal
surface case, the strain field induced by an ideal kinked vicinal
surface has not been calculated yet.

\section{Conclusion}

We have developed a general simple method to calculate the intensity
scattered by different kinds of surface morphologies that are not
perfectly ordered. The model based on the paracrystal framework
allows to recover previously published results on the scattering
from vicinal surfaces, two-level models and randomly stepped
surfaces. The most important result is the ability to characterize
two dimensional surface structures from nearest neighbor distance
distribution between steps or between kinks. The main hypothesis of
the model consists in neglecting the long-range correlations between
neighboring steps (or kinks) and, for the two dimensional models, in
assuming that the step edge shape and the terrace width are
independent. In the future, additional correlations should be
considered, e.g. in the field of thin film growth between the size
of neighboring terraces as highlighted first by M. C. Bartelt and J.
W. Evans \cite{bartelt93}. A step in this direction has been
proposed in the case of the two-level model (see Ref \cite{leroy04})
which should be extended to more complicated surface morphologies.

\section*{Acknowledgments}

We wish to acknowledge Bernard Croset (INSP, Paris) and Pierre
Muller (CRMCN, Marseilles) for fruitful discussions.
%\end{linenumbers}

%%%%%%%%%%%%%%%%%%%%%%%%%%%%%%%%%%%%%%%%%%%%%%%%%%%%%%%%%%%%%%%%%%%%%%
% --------------------------------------------------------------------
%               Appendices
% --------------------------------------------------------------------
%%%%%%%%%%%%%%%%%%%%%%%%%%%%%%%%%%%%%%%%%%%%%%%%%%%%%%%%%%%%%%%%%%%%%%

\appendix

\section*{Appendix A: Intensity scattered by a vicinal surface with a step height distribution}

\begin{center}
\begin{figure}
\includegraphics[width=12cm]{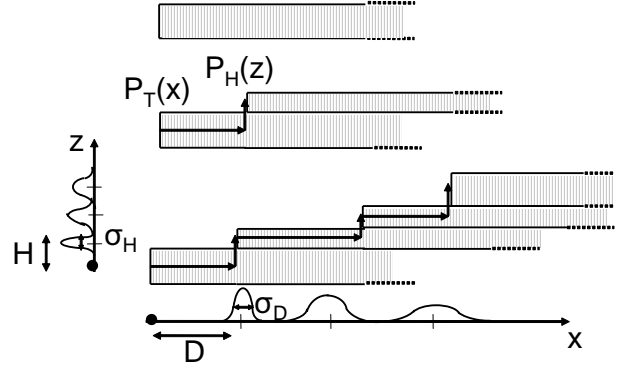}
\caption{Vicinal surface model assuming a terrace width distribution
($P_{T}$) as well as a step height distribution ($P_{H}$). D and H
are the mean terrace width and the mean step height and $\sigma
_{T}$ and $\sigma_{H}$ are the FWHM of the probability laws. The
FWHM of the probability to find two steps separated by a certain
distance increases with this distance as schematically illustrated
on the $x$-axis (same for the height, $z$-axis).}
\label{Vicinalestepheightdistribution}
\end{figure}
\end{center}

In this section, the intensity scattered by a vicinal surface with
both step height $P_{H}$ and terrace width $P_{D}$ distributions is
calculated (Fig~ \ref{Vicinalestepheightdistribution}).

The considered scattering object is a packing of H crystallographic
planes, semi-infinite and parallel to the terrace. H is distributed
according to the probability law $P_{H}$.

The auto-correlation function of the vicinal surface is (only the
contribution of steps at $x>0$ is given):
\begin{eqnarray}
g_{+}(x,z)&=& \iiiint_{-\infty}^{\infty} [F_{0}(x,z)\otimes F_{0}(x,z)\otimes \delta(z)\otimes \delta(x) + \nonumber \\
&+& F_{0}(x,z)\otimes F_{1}(x,z)\otimes
\delta(z-\frac{H_{0}+H_{1}}{2})\otimes P_{D_{0}}(x) + \nonumber \\
&+& F_{0}(x,z)\otimes F_{2}(x,z)\otimes
\delta(z-\frac{H_{0}+H_{2}}{2})\otimes P_{D_{0}}(x)\otimes [P_{H_{1}}(x)\otimes P_{D_{1}}(x)] +... \nonumber \\
&+& F_{0}(x,z)\otimes F_{n}(x,z)\otimes
\delta(z-\frac{H_{0}+H_{n}}{2})\otimes P_{D_{0}}(x)\otimes
[P_{H}(x)\otimes P_{D}(x)]^{\otimes^{n-1}}+... \nonumber \\
&+...& ]\ud P_{D_{0}}...\ud P_{D_{n-1}} \ud P_{H_{0}}...\ud
P_{H_{n}}...
\end{eqnarray}

where $F_{n}(x,z)$ is the shape of the $n^{th}$ object, i.e.
$F_{n}(x,z)=1$ at the positions of the unit cells otherwise it is
zero.

The intensity is obtained by Fourier transform of the
auto-correlation function :

\begin{eqnarray}
&&I(q_{x},q_{z})=FT\left \{ g(x,z)\right \} \nonumber \\
&=& \int_{H_{0}} \mid \widetilde{F}_{0}\mid ^2+ \nonumber \\
&+&\Big[\iiint_{H_{0}, H_{1}, D_{0}} \widetilde{F}_{0}(q_{x},q_{z})\widetilde{F}_{1}^*(q_{x},q_{z})e^{i(q_{z}\frac{(H_{0}+H_{1})c}{2})}e^{iq_{x}D_{0}a}+\nonumber \\
&+& \iiiint_{H_{0},H_{1},H_{2},D_{0},D_{1}} \widetilde{F}_{0}(q_{x},q_{z})\widetilde{F}_{2}^*(q_{x},q_{z})e^{i(q_{z}\frac{(H_{0}+H_{2})c}{2})}e^{iq_{z}D_{0}a}e^{i(q_{x}D_{1}a+q_{z}H_{1}c)}+...\nonumber \\
&...+& \iiiint_{H_{0},...,H_{n},D_{0},...,D_{n-1}} \widetilde{F}_{0}(q_{x},q_{z})\widetilde{F}_{n}^*(q_{x},q_{z})e^{i(q_{z}\frac{(H_{0}+H_{n})c}{2})}e^{iq_{z}D_{0}a}e^{i(q_{x}(D_{1}+...+D_{n-1})a+q_{z}(H_{1}+...+H_{n-1})c)}+...\nonumber \\
&...+& c.c. \Big]\ud P_{D_{0}}...\ud P_{D_{n-1}}\ud P_{H_{0}}...\ud
P_{H_{n}}...
\end{eqnarray}
As the structure factor of the object is:
\begin{eqnarray}
\widetilde{F}_{n}(q_{x},q_{z})&=&
\widetilde{F}_{cell}(q_{x},q_{z})\frac{e^{-iq_{x}a/2}}{-2i\sin
{(q_{x} a/2)}}\frac{\sin {(q_{z} H_{n}c/2)}}{\sin {(q_{z} c/2)}}\nonumber \\
&=& \widetilde{F}_{cell}(q_{x},q_{z})\frac{e^{-iq_{x}a/2}}{-2i\sin
{(q_{x} a/2)}}\frac{e^{i(q_{z} H_{n}c/2)}-e^{-i(q_{z}
H_{n}c/2)}}{2i\sin
{(q_{z} c/2)}}\nonumber \\
\end{eqnarray}
Therefore the scattered intensity reads:
\begin{eqnarray}
I(q_{x},q_{z}) &=& \left [\frac{\mid
\widetilde{F}_{cell}(q_{x},q_{z})\mid}{(2\sin
{(q_{x} a/2)}2\sin {(q_{z} c/2)})}\right ]^2 \times\nonumber \\
&\times& 2Re \left
\{1-\widetilde{P}_{H}(q_{z}c)+\frac{(\widetilde{P}_{H}(q_{z}c)-1)(1-\widetilde{P}_{H}(q_{z}c))\widetilde{P}_{D}(q_{x}a)}{1-\widetilde{P}_{H}(q_{z}c)\widetilde{P}_{D}(q_{x}a)}\right
\}\nonumber \\
&=&\left [\frac{\mid \widetilde{F}_{cell}(q_{x},q_{z})\mid}{(2\sin
{(q_{x} a/2)}2\sin {(q_{z} c/2)})}\right ]^2 Re \left
\{2\frac{(1-\widetilde{P}_{H}(q_{z}c))(1-\widetilde{P}_{D}(q_{x}a))}{1-\widetilde{P}_{H}(q_{z}c)\widetilde{P}_{D}(q_{x}a)}\right
\}\nonumber \\
\end{eqnarray}

which is the expected result.

\section*{Appendix B: Discrete distributions and characteristic functions}

Some examples of discrete distributions are given as well as their
characteristic functions :

\noindent The Binomial distribution :
\begin{eqnarray}
P_{p, N}(n)&=&\mathrm{C}_N^n p^n (1-p)^{N-n} \nonumber \\
\widetilde{P}_{p,N}(q)&=&(1-p+pe^{iq})^N  \nonumber
\end{eqnarray}
\noindent The Poisson distribution :
\begin{eqnarray}
P_{\nu}(n)&=&\frac{\nu^{n}}{n!} \nonumber \\
\widetilde{P}_{\nu}(q)&=&e^{\nu(e^{iq}-1)} \nonumber
\end{eqnarray}
\noindent The Geometric distribution :
\begin{eqnarray}
P_{p}(n)&=&p(1-p)^{n} \nonumber \\
\widetilde{P}_{p}(q)&=&\frac{p}{1-(1-p)e^{iq}} \nonumber
\end{eqnarray}
For completeness, more complex distributions can be considered
such as the hypergeometric distribution or the negative binomial
distribution.

%%%%%%%%%%%%%%%%%%%%%%%%%%%%%%%%%%%%%%%%%%%%%%%%%%%%%%%%%%%%%%%%%%%%%%
% --------------------------------------------------------------------
%               Biblio
% --------------------------------------------------------------------
%%%%%%%%%%%%%%%%%%%%%%%%%%%%%%%%%%%%%%%%%%%%%%%%%%%%%%%%%%%%%%%%%%%%%%
%\section{}
%\label{}

% \appendix

% \section{}
% \label{}

% Bibliographic references with the natbib package:
% Parenthetical: \citep{Bai92} produces (Bailyn 1992).
% Textual: \citet{Bai95} produces Bailyn et al. (1995).
% An affix and part of a reference:
%   \citep[e.g.][Ch. 2]{Bar76}
%   produces (e.g. Barnes et al. 1976, Ch. 2).

%\begin{thebibliography}{200}
%\bibliography{biblioFred-03-05-2007}
%\bibliography{biblioFred1}
% \bibitem[Names(Year)]{label} or \bibitem[Names(Year)Long names]{label}.
% (\harvarditem{Name}{Year}{label} is also supported.)
%\bibitem[]{}
%\end{thebibliography}

%%%%%%%%%%%%%%%%%%%%%%%%%%%%%%%%%%%%%%%%%%%%%%%%%%%%%%%%%%%%%%%%%%%%%%
% --------------------------------------------------------------------
%               FIGURE CAPTIONS AND FIGURES
% --------------------------------------------------------------------
%%%%%%%%%%%%%%%%%%%%%%%%%%%%%%%%%%%%%%%%%%%%%%%%%%%%%%%%%%%%%%%%%%%%%%
\newpage

\newpage

\newpage

\newpage

\newpage

\newpage

\newpage

\newpage

\newpage

\newpage

\newpage

\newpage

\newpage

\newpage

\newpage

\newpage

\newpage

\newpage

\end{document}